\documentclass[11pt]{article}
\usepackage{amsmath,amssymb,amsfonts, amsbsy, epsfig, psfrag,epsf}
\usepackage{graphicx}
\usepackage{enumerate}
\usepackage[authoryear]{natbib}
\usepackage{url} 
\usepackage{rotating}
\usepackage{boxedminipage}
\usepackage{algorithm}
\usepackage{subfigure}
\usepackage{multirow}
\usepackage{enumerate}
\usepackage{array}
\usepackage{xcolor}
\usepackage{algpseudocode}
\usepackage{pifont}
\usepackage{fullpage}
\usepackage{setspace}

\renewcommand{\baselinestretch}{1.6}

\algnewcommand{\Inputs}[1]{%
  \State \textbf{Inputs:}
  \Statex \hspace*{\algorithmicindent}\parbox[t]{.8\linewidth}{\raggedright #1}
}
\algnewcommand{\Initialize}[1]{%
  \State \textbf{Initialize:}
  \Statex \hspace*{\algorithmicindent}\parbox[t]{.8\linewidth}{\raggedright #1}
}

\algnewcommand{\Outputs}[1]{%
  \State \textbf{Outputs:}
  \Statex \hspace*{\algorithmicindent}\parbox[t]{.8\linewidth}{\raggedright #1}
}

\newcommand{\RR}{ {\mathbb{R}}  }
\newcommand\ind[1]{\mathbf{1}_{\{#1\}}}
\newcommand{\MA}{ {\mathcal{A}}}
\newcommand{\MF}{ {\mathcal{F}}}

\newcommand{\blind}{1}

\newtheorem{theorem}{Theorem}

\newtheorem{lemma}[theorem]{Lemma}

\newtheorem{proposition}[theorem]{Proposition}
\newtheorem{remark}{Remark}

\newcommand{\argmin}{\mathop{\mathrm{arg\,min}{}}}

\let\hat\widehat
\let\tilde\widetilde

\newcommand{\hth}{{\widehat \theta}}

\renewcommand{\P}{{\mathbb P}}
\newcommand{\E}{{\mathbb E}}

\definecolor{DSgray}{cmyk}{0,1,0,0}

\begin{document}

\def\spacingset#1{\renewcommand{\baselinestretch}%
{#1}} \spacingset{1.4}


\if1\blind
{
  \title{\bf Optimal Stopping and Worker Selection in Crowdsourcing: an Adaptive Sequential Probability Ratio Test Framework}
  \author{Xiaoou Li\\
  School of Statistics, University of Minnesota\\
  Yunxiao Chen \hspace{.2cm}\\
Department of Psychology, Emory University\\
    Xi Chen \\ 
    Stern School of Business, New York University \\
    Jingchen Liu, Zhiliang Ying \\
     Department of Statistics, Columbia University
    }
  \date{}
  \maketitle
} \fi


\if0\blind
{
  \bigskip
  \bigskip
  \bigskip
  \begin{center}
    {\LARGE\bf Optimal Stopping and Worker Selection in Crowdsourcing: an Adaptive Sequential Probability Ratio Test Framework}
\end{center}
  \medskip
} \fi

\bigskip
\begin{abstract}
In this paper, we aim at solving a class of multiple testing problems under the Bayesian sequential decision framework. Our motivating application comes from  binary labeling tasks in crowdsourcing, where the requestor needs to simultaneously decide which worker to choose to provide the label and when to stop collecting labels under a certain budget constraint. We start with the binary hypothesis testing problem to determine the true label of  a single object, and provide an optimal solution by casting it under the adaptive sequential probability ratio test (Ada-SPRT) framework.
We characterize the structure of the optimal solution, i.e., optimal adaptive sequential design, which minimizes the Bayes risk through log-likelihood ratio statistic. We also develop a dynamic programming algorithm that can efficiently approximate the optimal solution.  For the multiple testing problem, we further propose to adopt an empirical Bayes approach for estimating class priors and show that our method has an averaged loss that converges to the minimal Bayes risk under the true model. The experiments on both simulated and real data show the robustness of our method and its superiority in labeling accuracy as compared to several other recently proposed approaches.
\end{abstract}

\noindent%


\section{Introduction}

Over the past ten years, crowdsourcing has become an efficient and economical approach to obtaining labels for tasks that are difficult for computers but easy for humans. For example, the requestor can post a large amount of images on a popular crowdsourcing platform (e.g., Amazon Mechanical Turk) and ask a crowd of  workers to tag each picture as a portrait or a landscape with a small amount of payment to each label.

Despite its efficiency and immediate availability, the labels generated by non-expert crowd workers are quite noisy. As a remedy, most requestors resort to repetitive labeling for each object (e.g., an image), i.e., collecting multiple labels from different workers for an object. Then, the requestor aggregates the collected labels to infer the true label of each object. Generally, more labels for an object will lead to higher accuracy of the inferred true label. However, each label comes with a fixed amount of cost:  the requestor has to pay a pre-specified monetary cost for each obtained label, regardless of its correctness. Therefore, when using the crowdsourcing service for large-scale labeling tasks, a requestor usually faces two challenges:
\begin{enumerate}
  \item The requestor needs to carefully balance between the labeling accuracy and the cost of collecting labels. Basically, for each object, the requestor needs to decide when to stop collecting the next label based on the current information.
  \item The crowd workers have different levels of quality/reliablity. The requestor needs to  adaptively determine the next worker to label the object based on the current information.
\end{enumerate}

To address these challenges, we cast the problem into a general multiple testing problem in a sequential analysis framework. In particular, we study the most popular crowdsourcing task, binary labeling tasks, e.g., categorization of an image as a portrait or a landscape or a website as porn or not.  We assume that there are $K$ objects and for each object, we are interested in testing whether its true label (denoted by  $\theta_k \in \{0,1\}$) belongs to class zero or one.  More specifically, this problem can be formulated into a $K$ hypotheses testing problem
\begin{equation}\label{eq:multi}
H_{k0}: \theta_k=0 \quad \mbox{ against }  \quad  H_{k1}:\theta_k=1, \quad \mbox{ for } \; k=1,2,...,K.
\end{equation}
Since the true classes of objects  might be highly unbalanced, it is natural to assume that there is a prior $\pi_1$ (and $\pi_0=1-\pi_1$) such that
\begin{equation}\label{eq:class_prior}
 \pi_0=\P(\theta_k=0) \quad  \text{and} \quad  \pi_1=\P(\theta_k=1), \quad \mbox{ for } \; k =1,\ldots, K.
\end{equation}
The parameter $\pi_1$ models the unbalancedness between two classes, which is unknown to the algorithm.  To study this problem, let us first assume that the $\pi_1$ is known and consider the following hypothesis testing problem:
\begin{equation}\label{eq:single}
H_0: \theta=0 \quad \mbox{ against } \quad  H_1: \theta=1.
\end{equation}
To solve this problem, we propose an adaptive sequential probability ratio test (Ada-SPRT) in the Bayesian sequential analysis framework. We first formulate a {risk} function which is the expected probability of making wrong decision plus expected labeling cost. There are three components that we need to optimize over in this Ada-SPRT:
\begin{enumerate}
\item Stopping time: The first question is when to stop collecting more data/labels under a certain budget constraint (e.g., given a pre-specified maximum number of labels that can be collected). The early stopping is important for cost-effective crowdsourcing since a requestor should stop if the labels collected so far have already reached good consensus in order to avoid unnecessary cost.
\item Adaptive experimental selection rule: We assume that there are $M$ possible experiments (corresponding to heterogeneous workers), where different experiments lead to different likelihood functions for generating data/labels under true $\theta$. The key question is how to select the next experiment given the existing data.
\item Decision rule: Upon stopping, we need to make a decision on whether $H_0$  or $H_1$ is true.
\end{enumerate}
It is worthwhile to note that our Ada-SPRT can be viewed as an extension of the classical SPRT by \cite{Wald45,wald1948optimum}, which only optimizes the stopping time and decision rule but without the component of experiment selection.

In sequential analysis literature, \cite{chernoff1959sequential} and many follow-up works
have provided asymptotically optimal solutions for various sequential design problems (see Section \ref{sec:related} for more details). However, our problem has two challenges {beyond the classic asymptotic regime}:
\begin{enumerate}
  \item In crowdsourcing application, a requestor usually has limited budget (e.g., at most 10 labels for each object), which translates into an upper bound on the stopping time, that is, a truncation length. Under this constraint, the sample size cannot go to infinity and thus theory from asymptotically optimal experimental design will not hold any more.
  \item There is a class prior distribution $\pi_1$ in \eqref{eq:class_prior}, which needs to be estimated.
  However, to derive the traditional asymptotically optimal results, the effect of the prior probability distribution is usually  ignored as the expected sample size goes to infinity.

\end{enumerate}
To address these challenges and solve the general multiple testing problem in \eqref{eq:multi}, we propose an \emph{empirical Bayes approach} with a \emph{dynamic programming algorithm} to solve the single hypothesis testing problem in \eqref{eq:single} with a pre-specified truncation length $T$.  For a single truncated test, the sequential decision problem can be formulated into a Markov decision process (MDP), where the state space is characterized by log-likelihood ratio and the current sample size. To solve this MDP, we first provide a few structural results:
\begin{enumerate}
    \item The \emph{optimal stopping time} is a boundary hitting time based on the \emph{log-likelihood ratio}. The upper boundary curve is non-increasing with respect to (w.r.t.) the sample size $n=1, \ldots, T$ and lower boundary curve is non-decreasing w.r.t. the sample size $n$. 
  \item The \emph{optimal decision} for the true label is according to whether the \emph{log-likelihood ratio} hits the upper or lower boundary.
  \item The \emph{experiment/worker selection rule} is determined by the current \emph{log-likelihood ratio} and the sample size.
\end{enumerate}
Base on these structural results, we use a dynamic programming algorithm to solve the MDP.  We also characterize the relationship between the simpler non-truncated test (i.e., the truncation length $T = \infty$) and the truncated test and show that one can treat the non-truncated test as a limiting version of the truncated test as $T$ goes to infinity.

With the Ada-SPRT for solving \eqref{eq:single} in place, we solve the multiple testing problem in \eqref{eq:multi} using an empirical Bayes approach that estimates the class prior $\pi_1$. We prove
that as long as the class prior estimate is consistent, the averaged loss will converge to the minimal Bayes risk under the true model.  We further demonstrate its superior performance and robustness against different setups of true prior distribution (e.g., unbalanced class setting) using  empirical studies.

Finally, we would like to highlight that although we motivate our paper from a crowdsourcing application, our empirical Bayes with the Ada-SPRT approach is a general method for solving the multiple testing problem in \eqref{eq:multi}. The proposed method can be applied to a wide class of problems. 
For example, computerized mastery testing \citep{lewis1990using, ivan2004application, chang2005application, bartroff2008modern}, which is based on item response theory models \citep[e.g.][]{embretson2013item} and aims to classify examinees into ``mastery'' and ``non-mastery'' categories, has become an increasingly important testing mode in educational assessment.
The Ada-SPRT could be extended to provide an optimal adaptive mastery test design (in terms of Bayes risk) for each examinee. Such a test design will possess several advantages:  (1) items are selected sequentially, according to the current performance, (2) the test stops when enough information has been collected, and (3) classify the examinee as ``mastery" or ``non-mastery" based on the collected responses. 


The rest of the paper is organized  as follows. In Section \ref{sec:related}, we discuss related works in crowdsourcing, sequential analysis, and empirical Bayes literature. In Section \ref{sec:model}, we present the crowdsourcing model and the Bayesian decision framework, along with our risk function in the form of Bayes risk. In Section \ref{sec:ada-sprt}, we provide the structure of optimal adaptive sequential designs in terms of log-likelihood ratio, and further develop  numerical algorithms for optimal worker selection, stopping time and decision for both truncated and non-truncated tests. In Section \ref{sec:EB}, we extend the algorithm to multiple {testing} and present an empirical Bayes approach for estimating class priors. In Section \ref{sec:exp}, we demonstrate the performance of the proposed algorithm on both simulated and real crowdsourcing datasets, followed by conclusions in Section \ref{sec:conclusion}. All the proofs are provided in the supplementary material.


\section{Related Works}
\label{sec:related}

Crowdsourcing, as the most popular paradigm for effectively collecting labels at low cost,  has received {a great deal of}  attention from researchers in statistics and machine learning communities. Many works in this field are solving a static problem, i.e., inferring true labels and workers' quality parameters based on a static set of labels  (see, e.g., \cite{Vikas:10,Oh:12,Zhang14Spectral}). Most of the works are based on the so-called Dawid-Skene model \citep{Dawid:79} for modeling workers' quality. We shall adopt the Dawid-Skene model in our paper, which is also known as the two-coin model for binary labeling tasks. For adaptive worker selection problem, there are relatively fewer  results in the existing literature.  \cite{Oh:12} proposed to assign workers according to a random bipartite graph. However, such an approach fails to utilize the collected labels.  {\cite{Chen15Crowd}  considered the fixed budget problem and formulated the problem into a Bayesian MDP. They studied two greedy policies to approximately solve the MDP: (1) the knowledge gradient (KG) policy, which chooses the best experiment/action that maximizes the expected reward for the next stage; (2) the optimistic knowledge gradient (Opt-KG)  policy, which chooses the best action that maximizes the maximum of the reward for collecting a positive label and that for a negative label. We will compare with these two greedy policies in our experiments (see Section \ref{sec:real} for details). It is also worthwhile to note that instead of pre-fixing a total sampling budget as in \cite{Chen15Crowd}, our goal is to simultaneously conduct worker selection and make the optimal decision on stopping time.}

To achieve this goal, we formulate the problem into a Bayesian sequential testing problem and propose an adaptive sequential probability ratio test (Ada-SPRT) framework. Sequential testing, starting with the seminal works  of \cite{Wald45} and \cite{wald1948optimum} for testing two simple hypotheses, is one of the most classical and well-studied problems in sequential analysis.
We refer readers to the survey article \citep{Lai01} and books \citep{Siegmund85Seq, Tartakovsky14Seq} for a comprehensive review.
Sequential tests have received a wide range of applications in areas such as
industrial quality control,  design of clinical trials, finance, educational testing, etc \citep{ lai2004power, bartroff2008efficient, bartroff2013sequential, Lai01, bartroff2008modern, Tartakovsky14Seq}.
  The problem of sequential adaptive experiment selection was initially treated by \cite{chernoff1959sequential}, which considers a Bayes risk that is defined similarly to that in \cite{wald1948optimum}.
Another related work is  \cite{robbins1974sequential}, which presents Monte Carlo and theoretical analysis on several adaptive treatment selection rules in clinical trial, trying to reduce the expected number of observations made on the inferior treatment.  In addition, the current work is related to the multi-armed bandit problem \citep{robbins1985some}, which has been studied in many areas, such as clinical trials \citep{press2009bandit}, online advertising \citep{chakrabarti2009mortal,babaioff2009characterizing},  and portfolio design \citep{hoffman2011portfolio}.
The current work provides theoretical results in the sequential hypothesis testing framework that simultaneously considers the optimality of stopping, decision, and experiment selection.

Empirical Bayes method has recently gained prominence, in both theory and applications \citep[e.g.][]{jiang2009general,jiang2010empirical,koenker2014convex,brown2009nonparametric,Efron13EB}. We refer to \cite{zhang2003compound}, \cite{Efron13EB}, and the references therein for a comprehensive review.
In particular, \cite{karunamuni1988empirical} combines the empirical Bayes method and sequential analysis and provide theoretical analysis for the asymptotic behavior of a specific stopping rule. The current work extends this idea to the optimal design. To the authors' best knowledge, this is the first result encompassing empirical Bayes method, sequential analysis and experiment selection simultaneously.

To highlight our contribution, we compare our  results on adaptive sequential testing with the existing ones, which,  in general, fall into one of the three classes: 1)  sequential hypothesis testing with an adaptive sequential design  in an  asymptotic regime; 2) sequential hypothesis testing without an adaptive design  in a non-asymptotic regime; 3) sequential hypothesis testing with an adaptive design  in a non-asymptotic regime.
The comparisons and major differences between our work and the existing methods are summarized below.

\renewcommand\labelenumi{\theenumi)}
\begin{enumerate}
\item  The hypothesis testing with {a sequential design in an asymptotic regime} was first studied in \cite{chernoff1959sequential}, followed by a large body of literature including \cite{albert1961sequential,tsitovich1985sequential,naghshvar2013active,naghshvar2013sequentiality,bessler1960theory,Nitinawarat:15}. This line of research focuses on the behavior of sequential designs
when their expected sample sizes grow large.
Asymptotically optimal properties for different procedures are  derived. In particular, \cite{chernoff1959sequential,albert1961sequential,tsitovich1985sequential,naghshvar2013active,naghshvar2013sequentiality,bessler1960theory} derive asymptotically optimal results in terms of achieving the asymptotic lower bound of the Bayes risk as  cost $c\to 0$. \cite{Nitinawarat:15} derive asymptotically optimal result from a non-Bayesian point of view. {They show that as the error probabilities tend to zero, the expected sample size of their procedure achieves the asymptotic lower bound under each hypothesis.}

Motivated from the crowdsourcing application, we consider a different regime where the sample size is not allowed to go to infinity (fixed the cost $c$ and with a maximum test length constraint $T$). Thus, methods and techniques for the asymptotic regime are not applicable to our problem.

\item The sequential hypothesis testing in a non-asymptotic regime was first considered by \cite{wald1948optimum}, followed by a vast literature including \cite{wald1947sequensial,wald1950bayes,sobel1949sequential,arrow1949bayes,bussgang1955optimum,irle1984optimality,nikiforov1975sequential,bertsekas1978stochastic,Shiryaev:78}. Under the non-asymptotic regime, SPRT is shown to be  optimal  from a non-Bayesian point of view \citep{wald1948optimum}. The optimal truncated and non-truncated  Bayesian sequential tests have  been developed in \cite{arrow1949bayes}.

{Theorem~\ref{thm:truncate}  extends results for the optimal Bayesian sequential test in  \cite{arrow1949bayes} by incorporating an adaptive design (i.e., enabling the adaptive selection of the next experiment based on the current information).}

\item
The study of general stochastic control problem under the non-asymptotic regime dates back to \cite{bertsekas1978stochastic,bellman1957markovian,howard1970dynamic,Shiryaev:78}. 
Recent works, including \cite{bai2016line,naghshvar2010active}, establish theoretical  properties of the  optimal procedure for specific  hypothesis testing problems with experiment design, under the non-asymptotic regime. In particular, \cite{naghshvar2010active} consider {the problem of a single non-truncated sequential test with $M\geq 2$ hypotheses (among which only one holds true).
In this paper, we study a multiple testing problem and develop an empirical Bayes approach. For each single test with $M=2$ hypotheses, we provide a refined result on the continuation region either with or without a maximum test length constraint $T$.}

\label{paragraph:comparison}
\end{enumerate}
\label{paragraph:comparison_1}

\section{Model and Problem Setup}
\label{sec:model}

In this section, we first introduce the problem setup with full generality, followed by the specific application to crowdsourcing.  For a single object with  true label $\theta \in \{0, 1\}$, we are interested in the hypothesis testing problem in \eqref{eq:single}. Let  $X_1, X_2, \ldots $ be the  observed responses. The selection of $n$-th \emph{experiment} depends on all the previous responses. In particular, let $I=\{1,\ldots, M\}$ be the \emph{experiment pool}  and $\delta_n \in I$ be the selected $n$-th experiment, we have $\delta_n = j_n(X_1,...,X_{n-1})$, where the function $j_n(\cdot)$ is the experiment selection rule that needs to be learned.  Denote by $J$ the sequence of experiment selection rule $\{j_n:n=1,2,\ldots\}$.

Given $\theta$ and $\delta_n$, we denote the probability mass or density function of $X_n \in \mathbb{R}^d$ by $f_{\theta, \delta_n}$. {We make the  assumption that there exists at least one experiment $\delta\in I$ such that the  Kullback-Leibler divergence are bounded away from zero and infinity, i.e.,
$$
0<\E\left[\log\frac{f_{0,\delta}(X)}{f_{1,\delta}(X)}\Big|\theta=0\right]<\infty \mbox{ and } 0<\E\left[\log\frac{f_{1,\delta}(X)}{f_{0,\delta}(X)}\Big|\theta=1\right]<\infty.
$$}
{Here, $X$ is a generic notation for an observation with the probability mass or density function $f_{\theta,\delta}(x)$.}
Under this assumption, the model is identifiable and the standard SPRT has a finite expected sample size.
We would like to point out that our results are applicable to both continuous and discrete observations.

We further consider a random sample size denoted by $N$, that is, the test stops once sufficient observations has been collected. {We consider the case that there is a deterministic upper bound, or truncation length $T$, on the stopping time, that is, $N\leq T$.}
 Given all the responses and the stopping rule, one is able to decide whether to continue collecting at least one more response or to stop the test. Upon stopping, one is able to make a decision between $H_0$ and $H_1$. We denote by $D$ the decision rule, where $D=1$ represents $H_1$ is chosen while $D=0$ means $H_0$ is chosen.

The test procedure that has an experiment selection rule $J$, a stopping rule $N$, and a decision rule $D$ is called as an \emph{adaptive sequential design}. Our goal is to search for the optimal $J^{\dagger}$, $N^{\dagger}$ and $D^{\dagger}$ to minimize the composite risk of making a wrong test decision and the expected total labeling cost as defined below.

To define the risk,  we adopt the \emph{Bayesian decision framework.} In particular, we introduce the class prior
\begin{equation}\label{eq:class_prior}
 \pi_0=\P(\theta=0) \quad  \text{and} \quad  \pi_1=\P(\theta=1),
\end{equation}
with $\pi_0+ \pi_1=1$. We assume that  $\pi_1$ is known for the single hypothesis testing problem since it is impossible to estimate $\pi_1$ when there is only one object.  Let $c \in [0,1]$ be the \emph{relative cost} of collecting one response/label. The Bayes risk of an adaptive sequential test with experiment selection rule $J$, stopping time $N$ and decision rule $D$ is defined by \cite{wald1948optimum} as the expected probability of making wrong decision plus expected labeling cost,
\begin{align} \label{eq:bayesrisk}
      \mathbf{R}(J,N,D)  = &  \pi_0 \P(D=1|\theta=0)+\pi_1\P(D=0|\theta=1)  \\
                           & +c\{\pi_0 \E(N|\theta=0)+\pi_1\E(N|\theta=1)\}. \nonumber
\end{align}
We note that the \emph{relative cost} $c$, which is used to balance the trade-off between the labeling accuracy and labeling cost, needs to be set between zero and one. Since $\P(D=1|\theta=0) \leq 1$ and $\P(D=0|\theta=1) \leq 1$, to minimize the Bayes risk in \eqref{eq:bayesrisk}, one will not collect any label when $c>1$. In practice, the requestor usually chooses $c$ depending on the nature of labeling tasks (e.g., smaller $c$ for more challenging data to collect more labels) and the availability of the budget (e.g., a large $c$ for very limited amount of budget). \label{loc:c-clarify} We will demonstrate the effect of $c$ in our experiments in Section \ref{sec:exp}.

\begin{remark}\label{remark:weights}
Following the formulation in \cite{wald1948optimum}, we could consider a more general risk function by incorporating the weights $w_0,w_1>0$:
$$
R(J,N,D)= \pi_0 w_0 \mathbb{P}(D=1|\theta=0)+ \pi_1 w_1 \mathbb{P}(D=0|\theta=1) + c\{
\pi_0\mathbb{E}(N|\theta=0) + \pi_1 \mathbb{E}(N|\theta=1)
\}.
$$
Based on this more general formulation, similar analysis and algorithm could be developed with minor modification. In particular, in Theorem~\ref{thm:nontruncate}, we need to replace $\log(\frac{\pi_0}{\pi_1})$ by $\log(\frac{\pi_0 w_0}{\pi_1 w_1})$ and $c$ by $\frac{c}{\max(w_0,w_1)}$ respectively, in \eqref{eq:boundary-truncate2} and \eqref{eq:boundary-truncate1}. For the dynamic programming algorithm in Section~\ref{sec:algo}, we need to replace $\min\{\pi(\theta=0|l),\pi(\theta=1|l)\}$ by $\min\{w_0\pi(\theta=0|l),w_1\pi(\theta=1|l)\}$ in \eqref{eq:G_l_T}. We will also need to modify the definition of the averaged loss defined in \eqref{eq:D-compoundrisk} with the weights for Theorem~\ref{thm:EMB}.
\end{remark}

We denote by $\mathcal{A}^{T}$ the set of all  adaptive sequential designs $(J,N,D)$ such that the stopping time $N\leq T$.
We call the test procedure $(J^\dagger,N^\dagger,D^\dagger)$   an \emph{optimal test} among a class of adaptive sequential  testing procedure $\mathcal{A}^T$ (depending on the truncation length $T$)  if
\begin{equation}\label{eq:D-truncate}
   \mathbf{R}(J^\dagger,N^\dagger,D^\dagger)=\min_{(J,N,D)\in\mathcal{A}^T} \mathbf{R}(J,N,D).
\end{equation}

Now, for $K$ objects with true label $\theta_k \in \{0,1\}$ for $1 \leq k \leq K$, we consider the $K$ hypotheses testing problems with the unknown class prior $\pi_1$ in \eqref{eq:multi}. Let $D=\{D_k\}_{k=1}^K$ be the set of decisions and $N=\{N_k\}_{k=1}^K$ be the set of stopping times. The performance of the method is evaluated by the following \emph{averaged loss} defined over $K$ objects:
\begin{equation}\label{eq:D-compoundrisk}
L_K=\frac{1}{K}\sum_{k=1}^K \left[\ind{D_k\neq \theta_k}+cN_k\right].
\end{equation}
Our goal is to provide a \emph{consistent procedure}
 such that  $L_K$ converges to the minimal Bayes risk under the true model (i.e.,  $\min_{(J,N,D)\in\mathcal{A}^T} \mathbf{R}(J,N,D)$) in probability  as $K$ goes to infinity.

\subsection{Applications to crowdsourcing}
\label{sec:crowd}
Here, we briefly illustrate how this general sequential testing framework is connected to our motivating crowdsourcing application. We assume that there are $M$ workers (i.e., experiments) and we denote the set of workers by $I=\{1, \ldots, M\}$, which is our experiment pool.

For an object with the true label $\theta \in \{0,1\}$, let $\hth^i$ be the label provided by worker $i$, $i\in I$. The quality of worker $i$ is characterized by two quantities:
\begin{equation}\label{eq:two-coin}
\tau_{00}^i=\P(\hth^i=0 | \theta_i=0) \quad \text{ and }\quad \tau_{11}^i=\P(\hth^i=1 | \theta_i=1).
\end{equation}
In other words, $\tau_{00}^i$ is the probability that worker $i$ will provide the correct label to an object when the true label is zero and $\tau_{11}^i$ is that when the true label is one. This model is widely used in modeling crowd worker quality and is usually referred to as ``two-coin model" or Dawid-Skene model \citep{Dawid:79,Vikas:10,Zhang14Spectral}. For the ease of presentation, we  assume that $\tau_{00}^i$ and $\tau_{11}^i$ are given and will discuss, {in Section \ref{sec:EB}}, how to estimate these parameters in an online fashion as the labeling process goes  on.

The observed responses $X_n$ for $n=1,2, \ldots$ are the labels from the selected $n$-th worker $\delta_n$ according to the worker selection rule $j_n(X_1,...,X_{n-1})$. Under the two-coin model in \eqref{eq:two-coin}, each response takes the binary value, with  the following  probability mass function:
\begin{align}\label{eq:two_coin}
f_{\theta,\delta_n}(1)& =\P(X_n=1 | \delta_n, \theta)=\tau_{11}^{\delta_n} \ \ind{\theta=1}+(1-\tau_{00}^{\delta_n})\ \ind{\theta=0}, \\
f_{\theta,\delta_n}(0)&=\P(X_n=0 | \delta_n, \theta) = (1-\tau_{11}^{\delta_n}) \ \ind{\theta=1}+\tau_{00}^{\delta_n}\ \ind{\theta=0},\nonumber
\end{align}
where $\ind{\cdot}$ denotes the indicator function.


\section{Optimal Adaptive Sequential Probability Ratio Test}
\label{sec:ada-sprt}

In this section, we explore the structure of optimal adaptive sequential designs for the single hypothesis testing problem in \eqref{eq:single} and derive the dynamic programming algorithm to find optimal adaptive sequential designs.

\subsection{Structure of Optimal Adaptive Sequential Designs}
\label{sec:non_truncated}

We consider the class of truncated adaptive sequential  tests with the constraint that the sample size $N$ is no greater than a pre-fixed truncation length $T$.
{The optimization problem  \eqref{eq:D-truncate} is challenging because both the experiment selection and the stopping rule lie in infinite-dimensional function spaces.}
Our approach is to   make dimension reduction
by exploring the relationship between optimal adaptive sequential design with \emph{log-likelihood ratio statistics}.


In particular, under the optimal selection rule $J^\dagger=\{j^\dagger_1, j^\dagger_2, \ldots, \}$, the $n$-th selected experiment  (for $n \leq N^\dagger$)  is
\[
\delta_n^\dagger=j^\dagger_n(X_1, \ldots, X_{n-1}).
\]
The corresponding log-likelihood ratio statistic is defined by,
\begin{equation}\label{eq:D-loglik-truncate}
l^\dagger_n=\log\left(\frac{\prod_{i=1}^{n}f_{1,\delta^\dagger_i}(X_i)}{\prod_{i=1}^{n}f_{0,\delta^\dagger_i}(X_i)}\right), \quad \text{for} \; n=1,2, \ldots,
\end{equation}
where $f_{1, \delta^\dagger_i}(\cdot)$ and  $f_{0, \delta^\dagger_i}(\cdot)$ are the probability density/mass functions when $\theta=1$ and $\theta=0$ for the experiment $\delta_n^\dagger$, respectively. 
The next theorem
 characterizes the structure of the optimal adaptive  sequential design.

\begin{theorem}\label{thm:truncate}
Let $(J^{\dagger},N^{\dagger},D^{\dagger})$ be the optimal  adaptive  truncated sequential design as defined in \eqref{eq:D-truncate}. Then $(J^{\dagger},N^{\dagger},D^{\dagger})$  has the following properties.
\begin{enumerate}
\item[(i)]
The stopping time $N^{\dagger}$ is described through the hitting boundary of the log-likelihood ratio and the current sample size. In particular, there exist  two sequences of real values $A^{\dagger}(n)$ and $B^{\dagger}(n)$ for $1 \leq n \leq T$ such that
 \begin{align}
& \log\frac{\pi_0}{\pi_1}= A^{\dagger}(T)\leq A^{\dagger}(T-1)\leq ...\leq A^{\dagger}(1) \leq \log\frac{\pi_0 (1-c)}{\pi_1 c},\label{eq:boundary-truncate2} \\
& \log\frac{\pi_0 c}{\pi_1 (1-c)}\leq B^{\dagger}(1)\leq B^{\dagger}(2)\leq ...\leq B^{\dagger}(T)=\log\frac{\pi_0}{\pi_1}
\label{eq:boundary-truncate1},
\end{align}
 and the optimal stopping for the truncated test is determined by
\begin{equation}\label{eq:opt_stop_truncated}
N^{\dagger}=\inf\{n: l^{\dagger}_n\geq A^{\dagger}(n) \mbox{ or }l^{\dagger}_n \leq B^{\dagger}(n)\}.
\end{equation}
\item[(ii)]If $N^{\dagger}<T$, then the decision rule is
\[
D^{\dagger}=1  \; \text{ if }\; l^{\dagger}_{N^\dagger}\geq A^{\dagger}(N^{\dagger}) \quad \text{ and }\quad D^{\dagger}=0 \text{ if } \;  l^{\dagger}_{N^{\dagger}}\leq B^{\dagger}(N^{\dagger}).
\]
If $N^{\dagger}=T$ where $A^{\dagger}(T)= B^{\dagger}(T)$,  then
\[
D^{\dagger}=1 \; \text{ if  }\; l^{\dagger}_T\geq A^{\dagger}(T) \quad  \text{and} \quad  D^{\dagger}=0 \;  \text{ if } \; l^{\dagger}_T< B^{\dagger}(T).
\]
\item[(iii)] There exists an experiment selection function $j^{\dagger}:\mathbb{R}\times\{1,2,...\}\to I$ such that for $n=1,2,...,T$,
$$
\delta^{\dagger}_n= j^{\dagger}(l^{\dagger}_{n-1},n),
$$
where $\delta^{\dagger}_n$ is the $n$-th selected experiment under the optimal selection rule $J^{\dagger}$.
\end{enumerate}
\end{theorem}
\begin{remark}\label{remark:existence}
  According to Corollary 8.5.1 in \cite{bertsekas1978stochastic}, the optimal sequential adaptive design $(J^{\dagger},N^{\dagger},D^{\dagger})$ always exists (not necessarily unique) for the truncated test. We also note that the existence of the optimal design  for non-truncated problems when $T=\infty$ (see Proposition~\ref{thm:nontruncate} in  the later section) is guaranteed by Corollary 9.17.1 in \cite{bertsekas1978stochastic}.
\end{remark}

The proof of Theorem \ref{thm:truncate} is provided in the supplement material.
The statements (i) and (ii) are extensions of the seminal work of SPRT \citep{wald1948optimum} to the case of adaptive experiment selection. In contrast to the classical SPRT where the hitting boundaries are flat, the hitting boundaries for the truncated adaptive test include one non-increasing  curve (i.e., the upper boundary $A^{\dagger}(T)\leq A^{\dagger}(T-1)\leq ...\leq A^{\dagger}(1)$) and one non-decreasing curve (i.e., the lower boundary $B^{\dagger}(1)\leq B^{\dagger}(2)\leq ...\leq B^{\dagger}(T)$). Please see Figure~\ref{fig:boundary} for an illustration. Of note, since $A^{\dagger}(T)$ and $B^{\dagger}(T)$ take the same value $\log\frac{\pi_0}{\pi_1}$, the optimal stopping time $N^{\dagger}$  defined in \eqref{eq:opt_stop_truncated} automatically satisfies the constraint $N^{\dagger}\leq T$. The experiment selection  rule depends  on both the log-likelihood ratio statistic in \eqref{eq:D-loglik-truncate} and the current sample size.


\subsection{Dynamic Programming Algorithm}
\label{sec:algo}

Given the structure of optimal adaptive sequential design, we present a dynamic programming  algorithm for finding optimal experiment selection rule and hitting boundaries.

To describe the algorithm, we first introduce some necessary notations. Let $G(l,n) $ be the conditional risk associated with the log-likelihood ratio $l$ and the current sample size $n \in \{1, \ldots, T\}$. When the sample size $n$ reaches the truncation length $T$,  the testing procedure has to stop. For each $l$, we have
\begin{eqnarray}\label{eq:G_l_T}
  G(l,  T) = \min\{ \pi(\theta=0 | l) , \pi(\theta=1 | l)\} +  Tc,
\end{eqnarray}
where $\pi(\theta=0|l)$ and $\pi(\theta=1|l)$ are the posterior probabilities under the current log-likelihood ratio $l$ and $\min\{ \pi(\theta=0 | l) , \pi(\theta=1 | l)\}$ is the Bayes risk of making the wrong decision. The term $Tc$ is the cost of collecting $T$ responses. By the standard Bayesian decision theory (see e.g.,  \cite{Tartakovsky14Seq}, \S 3.2.2.)
\begin{equation*}
  \pi(\theta=0 | l) = \frac{\pi_0}{\pi_0+ \pi_1 e^{l}} \quad \text{and} \quad   \pi(\theta=1 | l) = \frac{\pi_1 e^{l}}{\pi_0+ \pi_1 e^{l}}.
\end{equation*}
Given the definition of $G(l,n)$, for any current sample size $n<T$ and log-likelihood ratio $l$, the optimal selection rule $j^{\dagger}(l,n+1)$ should choose the $(n+1)$-th experiment $\delta_{n+1} \in I$ to minimize the next stage expected conditional risk, i.e.,
\begin{eqnarray}
j^{\dagger}(l,n+1) = \argmin_{\delta\in I} \E_{l,\delta}G\left(l+\log\frac{f_{1,\delta}(X)}{f_{0,\delta}(X)}, \; n+1\right),
\end{eqnarray}
where the expectation is taken with respect to the next response $X$  when the next selected experiment is $\delta \in I$.

\label{paragraph:formulas}
{
As an illustration, we present an example of computing $\E_{l,\delta}G\left(l+\log\frac{f_{1,\delta}(X)}{f_{0,\delta}(X)}, \; n+1\right)$ when $n=T-1$ (corresponding to the first step in the dynamic programming algorithm).} In particular, we consider the two-coin model in \eqref{eq:two_coin} and consider the case for the $i$-th experiment. That is, $\delta=i$. Then, we have
  $$
  \log\frac{f_{1,i}(X)}{f_{0,i}(X)}=X\log\left(\frac{\tau_{11}^i}{1-\tau_{00}^i}\right)+(1-X)\log\left(\frac{1-\tau_{11}^i}{\tau_{00}^i}\right).
  $$
  To compute the conditional expectation of interest, we also need
  $$
  \P_{l,i}(X=1)=\pi(\theta=0|l)f_{0,i}(1)+ \pi(\theta=1|l)f_{1,i}(1) =  \frac{\pi_0}{\pi_0+ \pi_1 e^{l}} (1-\tau_{00}^i) + \frac{\pi_1 e^{l}}{\pi_0+ \pi_1 e^{l}}\tau_{11}^i.
  $$
  Combining the above two equations and \eqref{eq:G_l_T}, we have
  \begin{equation*}
      \begin{split}
   &\E_{l,\delta}G\left(l+\log\frac{f_{1,\delta}(X)}{f_{0,\delta}(X)}, \; n+1\right)\\
   =& \P_{l,i}(X=1) G\left(l+ \log\left(\frac{\tau_{11}^i}{1-\tau_{00}^i}\right),T\right) + (1-\P_{l,i}(X=1))G\left(l+ \log\left(\frac{1-\tau_{11}^i}{\tau_{00}^i}\right), T\right).
  \end{split}
  \end{equation*}
Now, we are ready to provide the recursive equation for $G(l,n)$, which is known as the Bellman equation in Markov decision process (see, e.g., \cite{Puterman:05,bertsekas1978stochastic}). In particular, under the current sample size $n$ and log-likelihood ratio $l$, the action for the next stage has two possible choices:
\begin{enumerate}
\item Stopping the testing procedure: the corresponding Bayes risk will be
    \[
     \min\{ \pi(\theta=0 | l) , \pi(\theta=1 | l)\} +  n c;
    \]
\item Collecting the next response from the experiment $j^{\dagger}(l, n+1)$ and the expected conditional risk becomes
    \[
        \E_{l,j^{\dagger}(l,n+1)}G\Big(l+\log\frac{f_{1,j^{\dagger}(l,n+1)}(X)}{f_{0,j^{\dagger}(l,n+1)}(X)}, \;n+1\Big).
    \]
\end{enumerate}
Combining these two cases, the requestor should choose the best possible action (either stop or continue) that leads to the minimum risk, resulting in the following recursive equation for $G(l,n)$,
\small
\begin{align*}
&  G(l,n) \\ = & \min\Big\{\E_{l,j^{\dagger}(l,n+1)}G\Big(l+\log\frac{f_{1,j^{\dagger}(l,n+1)}(X)}{f_{0,j^{\dagger}(l,n+1)}(X)},n+1\Big), \min\Big\{\frac{\pi_0}{\pi_0+\pi_1 e^l},\frac{\pi_1 e^l}{\pi_0+\pi_1 e^l}\Big\}+n c\Big\}.
\end{align*}
\normalsize
Finally, let $C(n)$ be the set of log-likelihood ratio at which the requestor should stop when the current sample size is $n$.
The upper hitting boundary $A^{\dagger}(n)$ and lower hitting boundary $B^{\dagger}(n)$ should then be the supremum and infimum of the log-likelihood ratio in $C(n)$. Given all the previous discussions, we present a dynamic programming algorithm for the truncated test in Algorithm \ref{alg:DP}.
\begin{remark}
  To implement Algorithm \ref{alg:DP} and solve for function $G(l,n), n=1,...,T$, discretization for $l$ and interpolation for $G(\cdot,n),n=1,...,T$ is necessary. That is, we approximate $G(\cdot,n),n=1,...,T$ with piecewise linear functions corresponding to the discretization over $l$. To justify this approximation, we notice that $G(l,n)$ is the minimum of finitely many continuous functions for $n=1,...,T$. Therefore, $G(l,n)$ is a continuous function in $l$ for $n=1,...,T$.
\end{remark}
\begin{remark}\label{remark:dp-complexity}
The computational complexity for the dynamic programming (DP) grows at the order
\[
O\left( T\times\text{discretization size the the likelihood ratio} \right),
\]
where $T$ is the maximum test length allowed. 
{
It is also worthwhile to note that the computation of the DP is done offline --- \emph{before} collecting any data and running the test.   Given the computational power nowadays, the offline computation is usually not considered as a computational burden.}
\end{remark}

\begin{algorithm}[!t]
\caption{Dynamic Programming for truncated Ada-SPRT}
\begin{algorithmic}[1]
\Inputs{$T$, $c$, $\pi_0$, $\pi_1$,  $\{f_{1,\delta}(\cdot)\}_{\delta \in I}$, $\{f_{0, \delta}(\cdot)\}_{\delta \in I}$   }
\Initialize{
$G(l,T)\gets \min\Big(\frac{\pi_0}{\pi_0+\pi_1 e^l},\frac{\pi_1 e^l}{\pi_0+\pi_1 e^l}\Big)+Tc,
$ for each $l$.\\
}
\For{$n$ = $T-1$ to 0 }
\State
$
j^{\dagger}(l,n+1)\gets\argmin_{\delta\in I} \E_{l,\delta}G\left(l+\log\frac{f_{1,\delta}(X)}{f_{0,\delta}(X)}, \; n+1\right)
$
\State
\[
G(l,n)\gets\min\Big\{\E_{l,j^{\dagger}(l,n+1)}G\left(l+\log\frac{f_{1,j^{\dagger}(l,n+1)}(X)}{f_{0,j^{\dagger}(l,n+1)}(X)},n+1\right), \min\Big\{\frac{\pi_0}{\pi_0+\pi_1 e^l},\frac{\pi_1 e^l}{\pi_0+\pi_1 e^l}\Big\}+nc\Big\}.
\]
\State
\[
C(n)\gets\Big\{l:\min\Big\{\frac{\pi_0}{\pi_0+\pi_1 e^l},\frac{\pi_1 e^l}{\pi_0+\pi_1 e^l}\Big\}+nc
\geq \E_{l,j^{\dagger}(l,n+1)}G\Big(l+\log\frac{f_{1,j^{\dagger}(l,n+1)}(X)}{f_{0,j^{\dagger}(l,n+1)}(X)}, \; n+1\Big)\Big \}.
\]
\State
$
A^{\dagger}(n)\gets
\arg\sup\{l: l\in C(n) \}.
$
\State
$
B^{\dagger}(n)\gets
\arg\inf\{l: l\in C(n) \}.
$
\EndFor
\Outputs{$j^{\dagger}$, $A^{\dagger}(n), B^{\dagger}(n)$ for $n=1,...,T$.}
\end{algorithmic}
\label{alg:DP}
\end{algorithm}

\subsection{Non-truncated Test}
\label{sec:non_truncate}
In this subsection, we investigate the relationship between the non-truncated ($T=\infty$) and the truncated test ($T<\infty$).
The structure of the optimal adaptive sequential design for a truncated test is simpler than that for the truncated test.
In particular, we extend the  result  
in \citet[Chapter~4.1, Lemma~1 and Theorem~1]{Shiryaev:78}, by adding the experiment selection component, and prove the following proposition on the structure of an optimal adaptive sequential design $(J^{*},N^{*},D^{*})$.  Let $\MA^*$ be the set of all the adaptive sequential designs such that both $\E(N|\theta=0)$ and $\E(N|\theta=1)$ are finite. We note that the assumptions $\E(N|\theta=0) < \infty$ and $\E(N|\theta=1 ) < \infty$ are commonly made in sequential analysis, e.g., \cite{wald1948optimum}.

\begin{proposition}\label{thm:nontruncate}
Let $(J^*,N^*,D^*)$ be an optimal  adaptive sequential design for a non-truncated test such that.
\begin{equation}\label{eq:D-nontruncate}
\mathbf{R}(J^{*},N^{*},D^{*})=\min_{(J,N,D)\in\mathcal{A}^*} \mathbf{R}(J,N,D).
\end{equation}
 Then $(J^*,N^*,D^*)$ has the following properties.
\begin{enumerate}
\item[(i)] The optimal stopping time $N^*$ is a boundary hitting time. That is, there exist real values $A^*$ and $B^*$ such that $B^*\leq A^*$ and
     $$
    N^*=\inf\{n: l^*_n\geq A^* \mbox{ or }l^*_n \leq B^*\}.
    $$

\item[(ii)] The optimal decision rule $D^*$ chooses between $H_0$ and $H_1$ according to whether the log-likelihood ratio statistic hits the upper or the lower boundary, i.e.,
    $$
        D^*=1 \; \text{ if }\; l^*_{N^*}\geq A^* \quad \text{ and } \quad D^*=0  \; \text{ if } \; l^*_{N^*}\leq B^*.
    $$
\item [(iii)] Each $j_n^*$ in the optimal experiment selection rule $J^*$ can be expressed as a single experiment selection function $j^*:\mathbb{R}\to I$ such that for any $n=1,2,...,N^*$,
    $$
    \delta^*_n=j^*(l^*_{n-1}).
    $$
\end{enumerate}
\end{proposition}
The proof of the Proposition \ref{thm:nontruncate} is provided in the supplement material.

\begin{remark}\label{remark:wald}
It was shown in \cite{wald1948optimum}  that if the stopping time is defined by the first passage time toward two flat boundaries, then the  expected sample size  is  minimized under the each hypothesis when the error probabilities are controlled. With adaptive experiment selection, such an optimal solution usually does not exist.
The main reason is that the best experiment selection rules are different under the null and alternative hypotheses, because the Kullback-Leibler information is not a symmetric function.  Thus, an informative experiment for one hypothesis may contain little information about the other. Consequently, the expected sample sizes under  both hypotheses may not be minimized simultaneously.
\end{remark}

\begin{remark}\label{remark:proof-MDP}
 The results in Proposition~\ref{thm:nontruncate}  can be proved using techniques from MDP from \cite{Shiryaev:78}, though the proof that we present in the supplement material follows the techniques developed in   \cite{wald1948optimum}.
\end{remark}

In contrast to the truncated case in Theorem \ref{thm:truncate}, the boundaries for non-truncated tests are flat. 
Moreover, the selection function $j^*$ is independent of the current sample size $n-1$ and depends on previous responses $X_1, \ldots, X_{n-1}$ only through the log-likelihood ratio statistic $l^*_{n-1}$.

The next theorem shows that in terms of the minimum Bayes risk,  the non-truncated test is a limiting version of the truncated test as $T \rightarrow \infty$.
\begin{theorem}\label{thm:truncate-approximate-nontruncate}
Let $\MA^T$ denote the set of all adaptive sequential designs $(J,N,D)$ such that $N\leq T$, and $\MA^*$ the set of all sequential adaptive designs that have finite expected sample size. Then,
$$
\lim_{T\to\infty}\min_{(J,N,D)\in \MA^T} \mathbf{R}(J,N,D)= \min_{(J,N,D)\in\MA^*}\mathbf{R}(J,N,D).
$$
\end{theorem}

The proof of Theorem \ref{thm:truncate-approximate-nontruncate} is provided in the supplement (see Section \ref{sec:proof_approx} in the supplement). 

\section{Multiple Hypotheses Testing and Empirical Bayes Approach}
\label{sec:EB}

So far, we have discussed optimal Ada-SPRT for a single object. Now we are ready to address our target problem in \eqref{eq:multi}, which is a $K$ hypotheses testing problem.  Let us recall the last paragraph in Section \ref{sec:model}, where  $D=\{D_k\}_{k=1}^K$ is the set of decisions and $N=\{N_k\}_{k=1}^K$ is the set of stopping times. The averaged loss $L_K$ is defined in \eqref{eq:D-compoundrisk}.

If the class prior $\pi_1$ is known, then, according to Theorem \ref{thm:truncate}, the optimal design that minimizes $\E L_K$ is that we run Algorithm \ref{alg:DP} independently for each object $k$ to obtain the optimal experiment selection rule (denoted by $j^{(k)}$) and boundaries or sequence of boundaries for the truncated case (denoted by $A^{(k)}$ and $B^{(k)}$). Given $j^{(k)}$, $A^{(k)}$ and $B^{(k)}$, the requestor collects labels according to the selection rule $j^{(k)}$ for each object $k$ and makes the decision $D_k$ according to the hitting boundary. Although such a procedure is easy to implement, the class prior $\pi_1$  and $\pi_0=1-\pi_1$ in \eqref{eq:class_prior} are unknown in many real-world applications. With multiple objects, one can estimate the class prior via the \emph{empirical Bayes} approach described as follows.

We assume that $\theta_k \in \{0,1\}$  for $k=1,\ldots, K$ are independently and identically distributed following the Bernoulli distribution with an \emph{unknown} parameter $\pi_1$. For each $k$, we estimate $\pi_1$ by some estimator $\hat \pi_1$ based on the collected responses for previous hypothesis $1,2, \ldots ,k-1$. In principle, any estimator can be applied to estimate $\pi_1$ and we adopt the maximum likelihood estimator.  Then, for the $k$-th hypothesis, we use Algorithm \ref{alg:DP} with the estimated parameters  $\hat \pi_1^{(k)}$, $\hat \pi_0^{(k)}=1-\hat \pi_1^{(k)}$ to solve for the experiment selection rule and stopping time for the $k$-th hypothesis. The algorithm is presented in Algorithm~\ref{alg:EMB}, where we initialize the estimate for $\pi_1$ to be 0.5 for simplicity.

\begin{algorithm}[!t]
\caption{Ada-SPRT for multiple objects using empirical Bayes method}
\label{alg:EMB}
\begin{algorithmic}[1]
\Inputs{$c$, $\{f_{1,\delta}(\cdot)\}_{\delta \in I}$, $\{f_{0, \delta}(\cdot)\}_{\delta \in I}$, $T$}
\Initialize{$\hat \pi_0^{(0)}=\hat\pi_1^{(0)}=0.5$}
\For{$k$=1 to $K$}
\State Run Algorithm~\ref{alg:DP} with Inputs $c$,  $\hat\pi_0^{(k-1)},\hat\pi_1^{(k-1)}$, $\{f_{1,\delta}(\cdot)\}_{\delta \in I}$, $\{f_{0, \delta}(\cdot)\}_{\delta \in I}$, and $T$. Obtain Outputs $A^{(k)},B^{(k)},j^{(k)}$.
\State Collect responses according to the experiment selection rule $j^{(k)}$ and obtain the decision $D_k$ according to the boundary hitting.
\State Update $\hat \pi_0^{(k)}$ and $\hat \pi_1^{(k)}$ with the newly collected responses.
\EndFor
\Outputs{Decision $D_k$ and sample size $N_k$ for each hypothesis $k=1, \ldots, K$.}
\end{algorithmic}
\end{algorithm}

As the number of hypotheses $K$ grows large, and the estimate $\hat \pi_1$ becomes very accurate, the resulting averaged loss $L_K$ in \eqref{eq:D-compoundrisk} will converge to the minimal Bayes risk corresponding to the true $\pi_1$. We characterize this asymptotic   result in the next theorem.

\begin{theorem}\label{thm:EMB}
Assume that $c<\pi_1<1-c$, and $\hat \pi_1\to \pi_1$ in probability as $K \rightarrow \infty$ and the sequential adaptive deign $D_k$ and $N_k$ are determined through the empirical Bayes procedure described in Algorithm~\ref{alg:EMB}, then
$$
 L_K\to \min_{(J,N,D)\in\mathcal{A}} \mathbf{R}(J,N,D) \mbox{ in probability} \mbox{ as }K\to\infty,
$$
where $ \mathbf{R}(J,N,D)$ is the  minimal Bayes risk of a single object defined in \eqref{eq:D-nontruncate}.
That is, the averaged loss $L_K$ in \eqref{eq:D-compoundrisk} converges to the minimal Bayes risk under the true model.
\end{theorem}

The proof of Theorem \ref{thm:EMB} is provided in the supplement (see Section \ref{sec:proof_EMB} in the supplement).   We also note that in Theorem~\ref{thm:EMB}, the assumption $c<\pi_1<1-c$ is a necessary condition for the optimal test procedure to be non-trivial, without which the optimal test will always stop with no sample.

In the sequential analysis literature, the likelihood functions $\{f_{1,\delta}(\cdot)\}_{\delta \in I}$ and $\{f_{0, \delta}(\cdot)\}_{\delta \in I}$ are typically assumed to be known.  However, in real crowdsourcing applications, it is quite often that no prior knowledge on workers' quality parameters  $\{\tau_{00}^i\}_{i \in I}$, $\{\tau_{11}^i\}_{i \in I}$ in \eqref{eq:two_coin} is available. Therefore, one cannot directly compute the likelihood ratio statistics in terms of $\{f_{1,\delta}(\cdot)\}_{\delta \in I}$ and $\{f_{0, \delta}(\cdot)\}_{\delta \in I}$. To address this issue, we propose to estimate the workers' quality parameters using a regularized maximum likelihood estimate under the two-coin model in \eqref{eq:two_coin} after finishing the labeling process for each object $k$. In particular, after each for-loop in Algorithm \ref{alg:EMB} (i.e., the labeling process and the decision for the $k$-th object has finished), we have collected all the responses $\{Z_{ji}\}$, where each $Z_{ji}$ is a binary label from worker $i \in I$ to the object $j\in \{1, \ldots, k\}$. A regularized minus log-likelihood is defined as follows,
\small
\begin{align}\label{eq:reg_likelihood}
& h_k\left(\pi_1, \{\tau_{00}^i\}_{i \in I},  \{\tau_{11}^i\}_{i \in I} \right)= \\
& -\sum_{1 \leq j \leq k} \log \left((1-\pi_1)\prod_{i} (\tau_{00}^i)^{1-Z_{ji}}(1-\tau_{00}^i)^{Z_{ji}} + \pi_1 \prod_{i }(1-\tau_{11}^i)^{1-Z_{ji}}(\tau_{11}^i)^{Z_{ji}}\right) \nonumber \\
&+ \sum_{i \in I }   \left((\alpha-1)\log (\tau_{00}^i) + (\beta-1)\log(1-\tau_{00}^i)  + (\alpha-1)\log (\tau_{11}^i) + (\beta-1)\log (1-\tau_{11}^i) \right). \nonumber
\end{align}

\normalsize
The regularization term comes from the Beta priors on $\tau_{00}^i$ and $\tau_{11}^i$ for each $i \in I$ with parameters $\alpha$ and $\beta$, which makes the estimation stable when a worker has only labeled a small number of objects.
We minimize $h_k\left(\pi_1, \{\tau_{00}^i\}_{i \in I},  \{\tau_{11}^i\}_{i \in I} \right)$ at the end of $k$-th iteration in Algorithm \ref{alg:EMB} using the expectation maximization (EM) algorithm \citep{Dempster1977}, which simultaneously provides the estimate of class prior $\pi_1$ (i.e., $\hat{\pi}_1^{(k)}$) and workers' quality parameters $\{\tau_{00}^i\}_{i \in I},  \{\tau_{11}^i\}_{i \in I}$ (see the details  in \cite{Dawid:79}). These estimates will be used to construct the optimal adaptive sequential designs for the next object $k+1$. After the decision for the $(k+1)$-th object has been made,  we re-optimize $h_{k+1} \left(\pi_1, \{\tau_{00}^i\}_{i \in I},  \{\tau_{11}^i\}_{i \in I} \right)$ using all the previously collected responses. We also adopt the estimate from the $k$-th iteration as the starting point (so-called warm-start) so that  the EM algorithm usually quickly converges in a few iterations.

\section{Experimental Results}
\label{sec:exp}

In this section, we conduct both simulated and real experiments to demonstrate the performance of the proposed Ada-SPRT algorithms.


\subsection{Simulated Experiments}

\subsubsection{Effect of Truncation Length $T$}
\label{sec:trun_effect}

\begin{figure}[!t]
\centering
\begin{minipage}[b]{0.35\textwidth}
\includegraphics[width=0.95\textwidth]{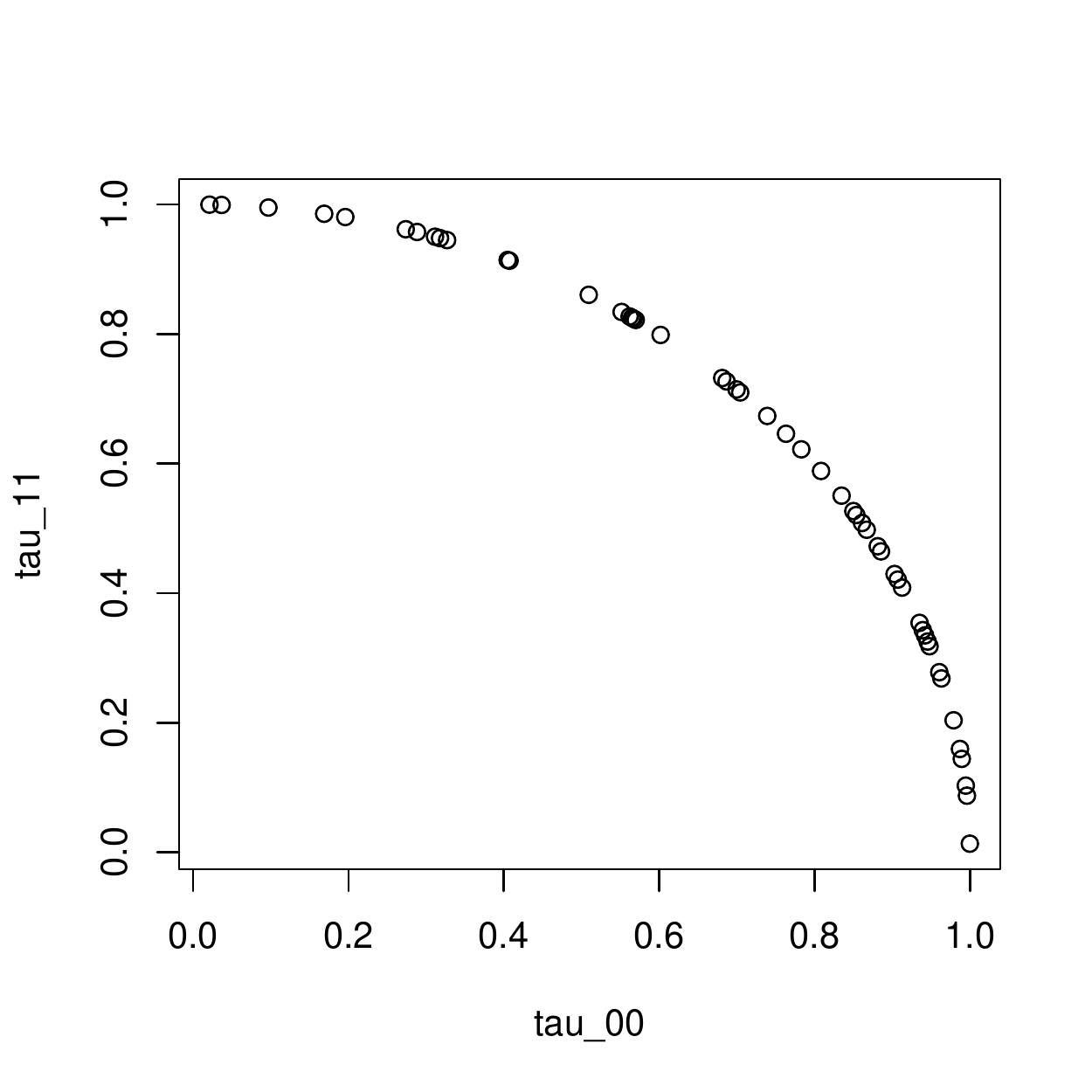}
\caption{The quality parameters for 50 simulated workers} 
\label{fig:delta}
\end{minipage}\hspace{0.3cm}
\begin{minipage}[b]{0.6\textwidth}
\includegraphics[width=0.95\textwidth]{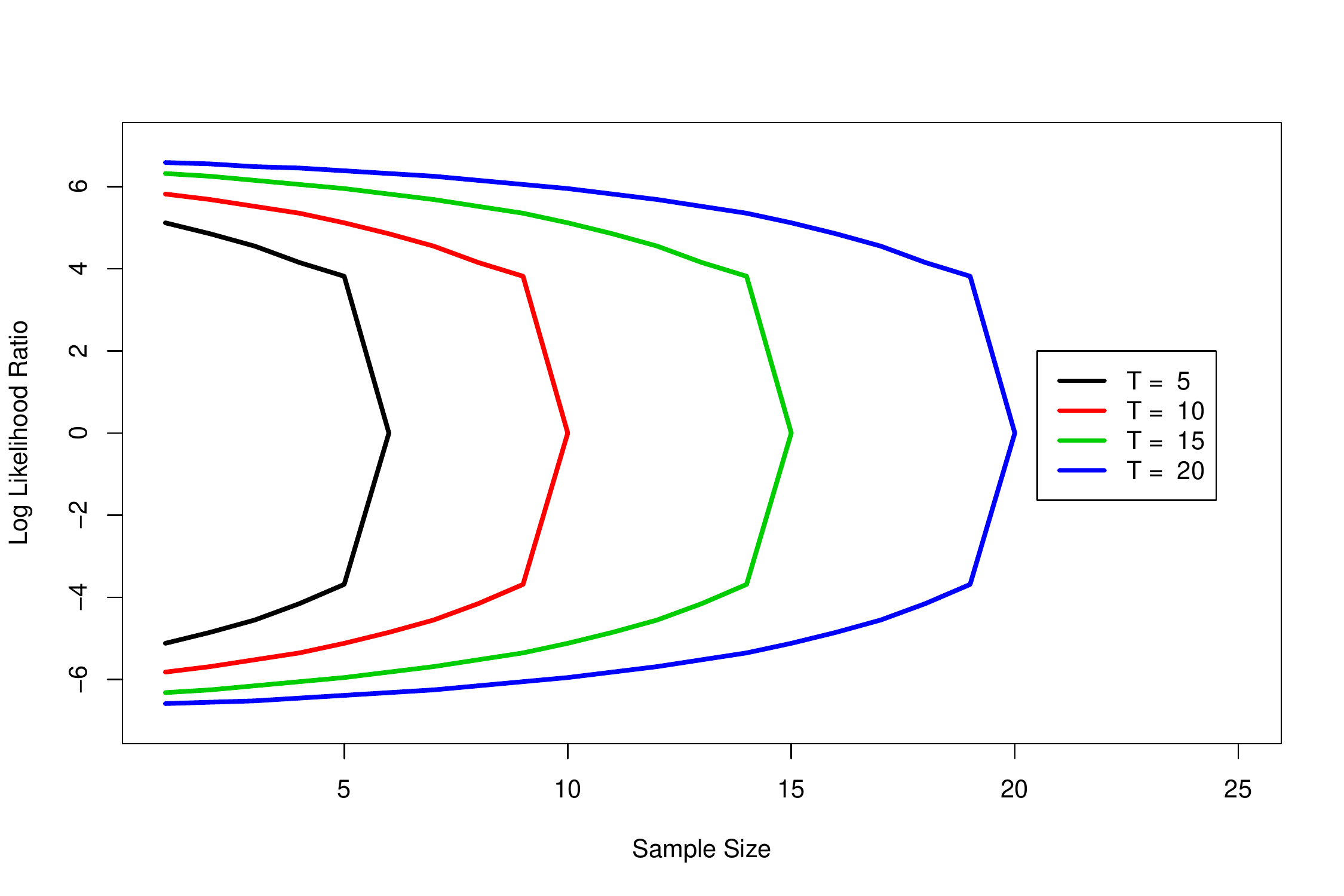}
\caption{Hitting boundaries for different truncation lengths.}
\label{fig:boundary}
\end{minipage}
\end{figure}


\begin{table}[!t]
\centering
\caption{Performance of Ada-SPRT for different truncation lengths.}
\vspace{0.2cm}
\begin{tabular}{lccccc}
\hline
 & $T = 5$& $T = 10$ & $T = 15$ & $T = 20$ \\
\hline
Stopping Time &5.000& 9.276& 12.301& 14.505\\
Accuracy &0.857 &0.926 & 0.961 & 0.977     \\
Loss &14.468& 7.672&  4.240&  2.630        \\
\hline
\end{tabular}
\label{table:length}
\end{table}

We first study the effect of the truncation length $T$ for a single hypothesis. We simulate $M=50$ workers with quality parameters   for  worker $i$:
\begin{align*}
&\gamma^i \sim \text{Uniform}(0, \frac{\pi}{2}),\\
&\tau^i_{00} = \sin(\gamma^i), \quad \tau^i_{11} = \cos(\gamma^i).
\end{align*}
A scatter plot of the generated $\tau^i_{00}$ for $1 \leq i \leq M$ is shown in Figure~\ref{fig:delta}. We generate 50 workers in this way such that no worker is dominantly worse than another.
That is, there does not exist a pair of workers $i$ and $i'$ such that $\tau^{i}_{00} < \tau^{i'}_{00}$ and $\tau^{i}_{11} < \tau^{i'}_{11}$.


We consider a single hypothesis testing problem (i.e., labeling for a single object) with the true label $\theta$ drawn from the Bernoulli distribution with $\pi_0=\pi_1=0.5$. In  this experiment, since our main goal is to investigate the effect of truncation length $T$, we assume true $\pi_1$ and workers' parameters are known for simplicity and set the parameter $c=2^{-12}$. We vary the truncation length $T = 5, 10, 15$, and $20$. For different truncation lengths, we plot the hitting boundaries in Figure \ref{fig:boundary}. As one can see, given any fixed truncation length $T$, for different sample sizes from 1 to $T$ (on the $x$-axis of Figure \ref{fig:boundary}), we have
\[
B^{\dagger}(1)\leq B^{\dagger}(2)\leq ...\leq B^{\dagger}(T)=\log\frac{\pi_0}{\pi_1}= 0= A^{\dagger}(T)\leq A^{\dagger}(T-1)\leq ...\leq A^{\dagger}(1).
\]
This observation is consistent with our result in Theorem \ref{thm:truncate}.  

Now for each truncation length $T$, we generate 50,000 independent replications and run Ada-SPRT for each replication. In Table~\ref{table:length}, we report  the \emph{average} of (1) the stopping time $N$, (2) the labeling accuracy $\ind{D=\theta}$, and (3) the loss  $\ind{D \neq \theta}+ c N$ over 50,000 replications. As can be seen from Table~\ref{table:length}, as the truncation length increases, both the stopping time and accuracy increase simultaneously. However, the average loss, which consists of labeling error and cost, decreases as $T$ goes larger.

\subsubsection{Comparison with the asymptotically optimal KL-information Approach}
\label{sec:KL}
\begin{figure}[!t]
\centering
\subfigure[$\pi_1=0.8$]{
  \includegraphics[width=0.31\textwidth]{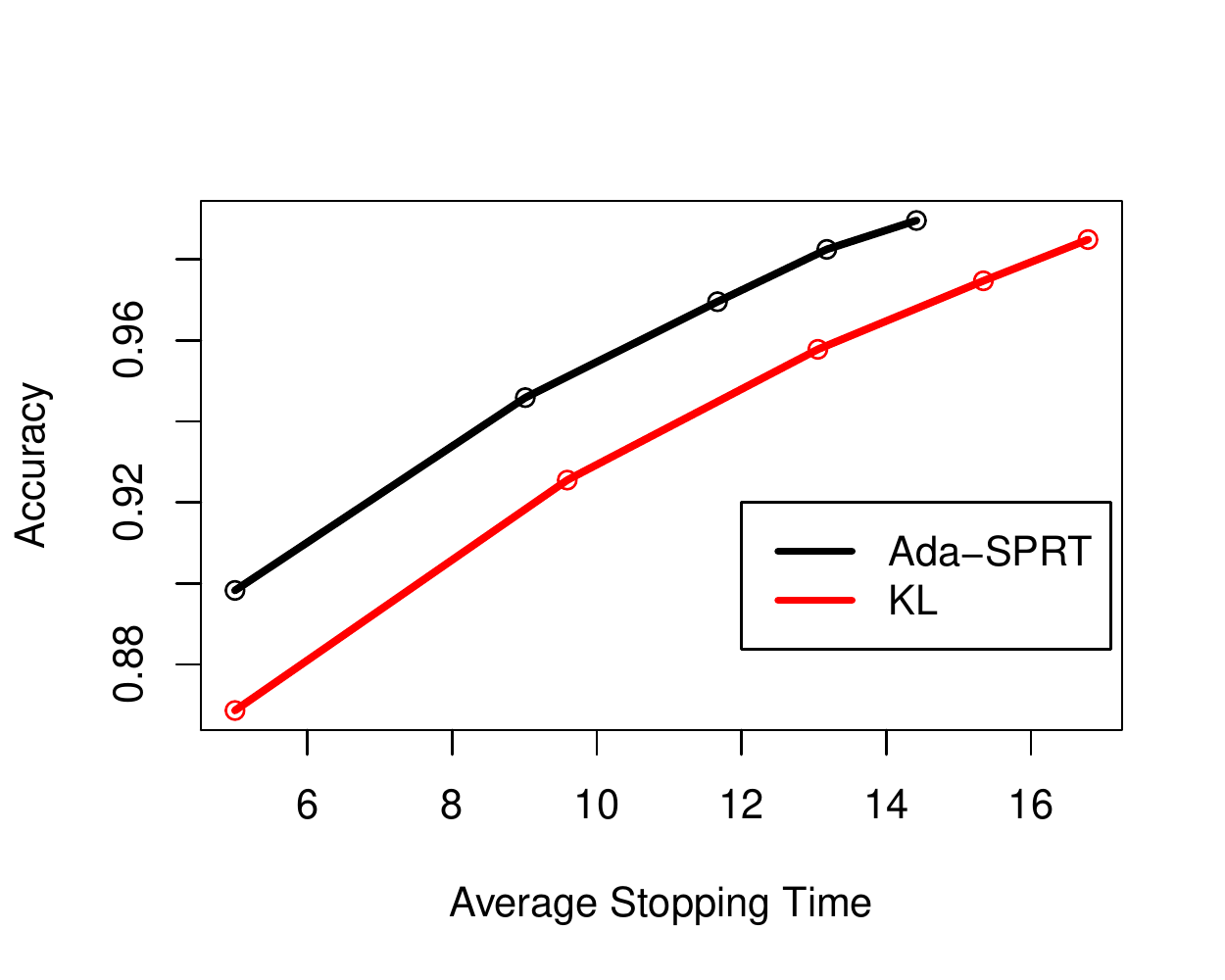}
} \vspace{0cm}
\subfigure[$\pi_1=0.65$]{
  \includegraphics[width=0.31\textwidth]{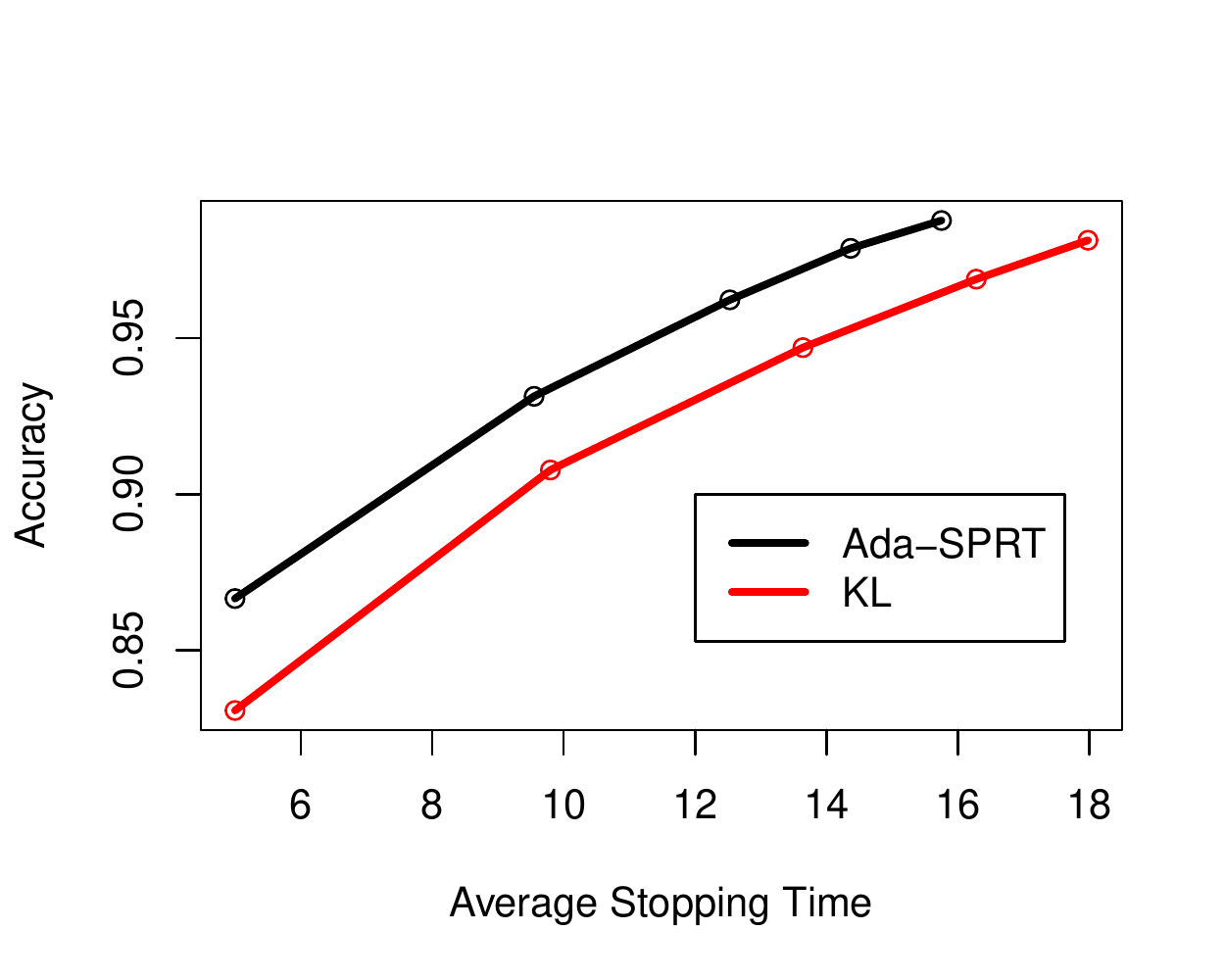}
} \vspace{0cm}
\subfigure[$\pi_1=0.5$]{
  \includegraphics[width=0.31\textwidth]{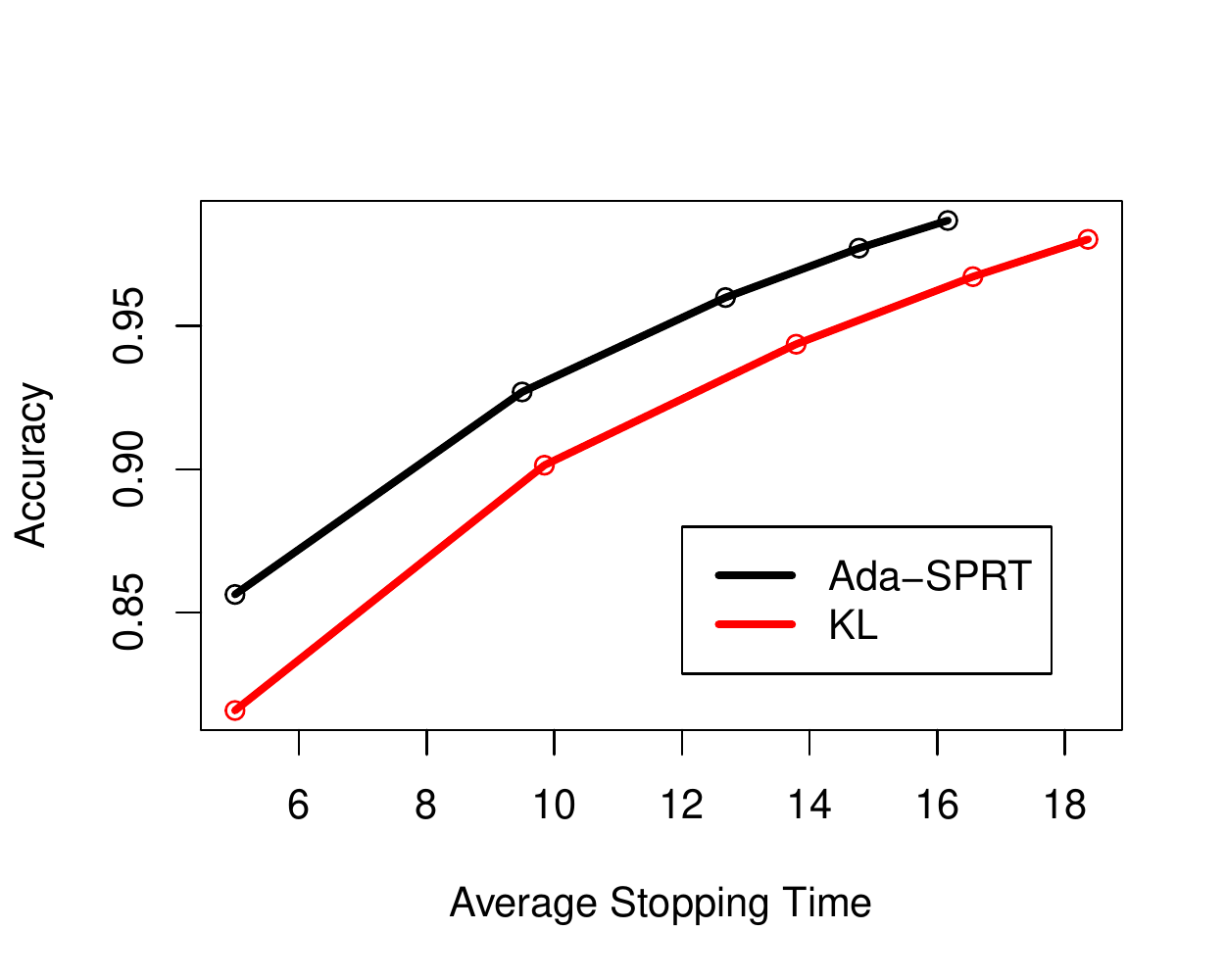}
}\vspace{0cm}
\caption{Comparison between the Ada-SPRT and KL approaches.}
\label{fig:KL}
\end{figure}

We compare the proposed Ada-SPRT procedure with an asymptotically optimal Kullback-Leibler (KL) approach from \cite{chernoff1959sequential}.
The worker selection rule of the KL approach is based on workers'
KL information, where the KL information for worker $\delta \in I$ given $\theta = 0$ and $\theta = 1$ is defined as
$$KL(0, \delta) = \E\left[\log\frac{f_{0,\delta}(X)}{f_{1,\delta}(X)}\big |\theta=0\right], ~~\mbox{and}~~ KL(1, \delta) =\E\left[\log\frac{f_{1,\delta}(X)}{f_{0,\delta}(X)}\big |\theta=1\right].$$
At  time $n$, let $\pi(\theta=0|l)$ and $\pi(\theta=1|l)$ be the posterior probabilities under the current log-likelihood ratio $l$. 
Then the worker selection rule of the KL approach is
\begin{equation*}
j(l, n) = \left\{
\begin{array}{ll}
  \arg\max_{\delta \in I} KL(0, \delta),  & \mbox{~if~~} \pi(\theta=0|l) > \pi(\theta=1|l),\\
  \arg\max_{\delta \in I} KL(1, \delta),  & \mbox{~otherwise}.
\end{array}\right.
\end{equation*}
That is, the worker with the largest KL information at the posterior mode of $\theta$ is selected.
In terms of the stopping rule, this KL approach adopts flat boundaries
$$A = -\log c + \log\left(\frac{\pi_0\max_{\delta\in I} KL(1, \delta)}{\pi_1}\right)\mbox{~~and~~} B = \log c + \log\left(\frac{\pi_0}{\pi_1\max_{\delta\in I} KL(0, \delta)}\right),$$
where the second terms in both $A$ and $B$ take the prior information and the worker pool quality into account.
The algorithm stops once the log-likelihood ratio $l$ crosses the boundaries, i.e., $l \geq A$ or $l \leq B$, or the sample size $n$ has reached the truncation length $T$. The decision is based on the posterior probabilities upon stopping, that is, $D = \arg\max_{d\in \{0, 1\}} \pi(\theta=d|l)$.

To compare the Ada-SPRT and KL approaches, the same worker pool in Section \ref{sec:trun_effect} is used. We consider three possible values of the class prior $\pi_1$: (1) $\pi_1=0.8$ (highly unbalanced class) (2) $\pi_1=0.65$ (moderately unbalanced class) (3)  $\pi_1=0.5$ (balanced class). We set $c = 2^{-12}$ and vary the truncation length $T = 5, 10, 15, 20, 25$.
For each $\pi_1$, $c$, and $T$, $500,000$ independent replications are generated. Results are summarized in Figure~\ref{fig:KL}, where for each choice of $\pi_1$, we report the average accuracy as a function of average stopping
time under varying truncation length $T$. According to Figure~\ref{fig:KL}, the proposed Ada-SPRT method performs substantially better than the KL procedure under a finite sample setting.



\subsubsection{Class Prior and Empirical Bayes Estimator}

\begin{figure}[!t]
\centering
\subfigure[$\pi_1=0.8$]{
  \includegraphics[width=0.3\textwidth]{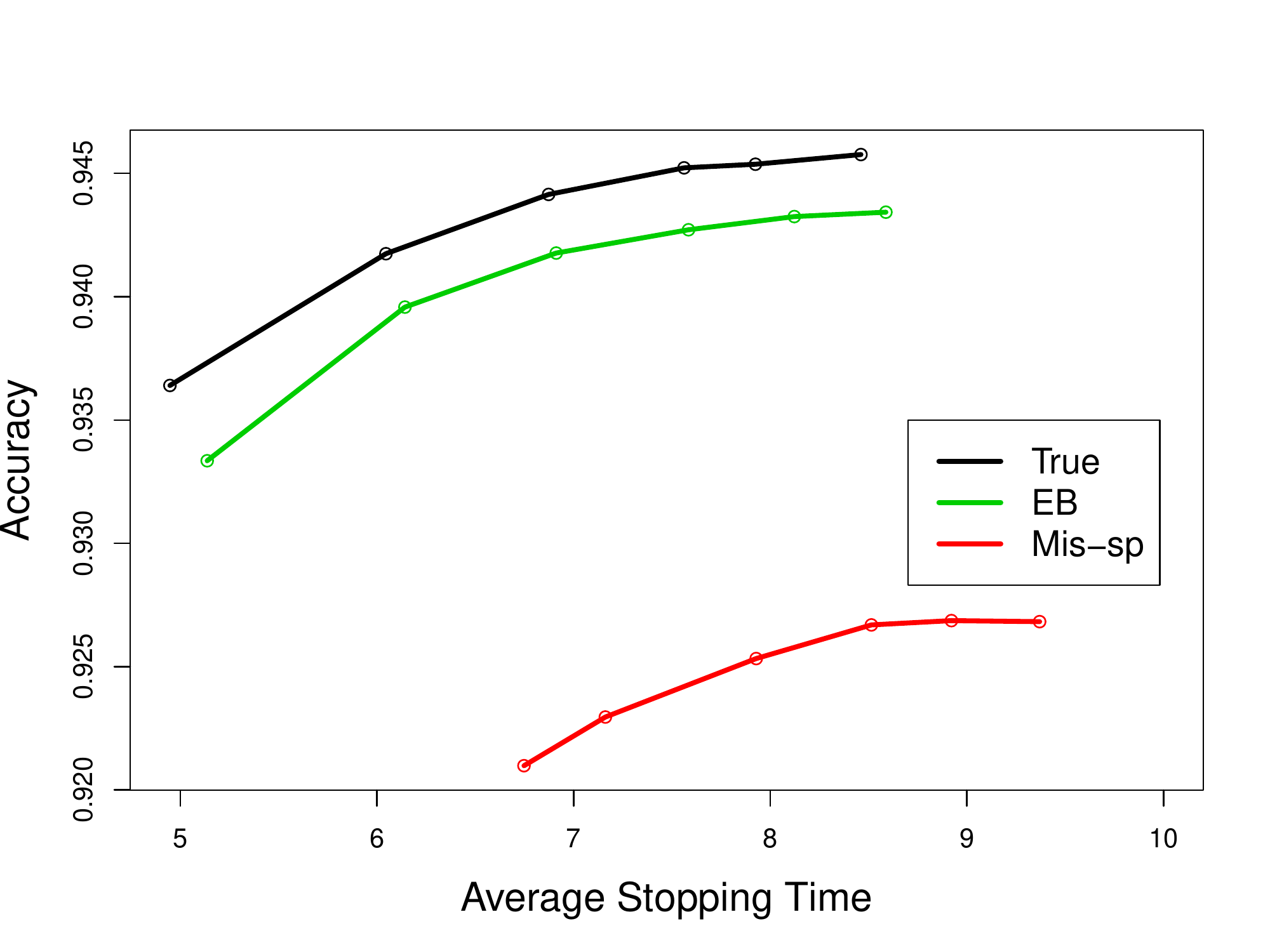}
} \vspace{0cm}
\subfigure[$\pi_1=0.65$]{
  \includegraphics[width=0.3\textwidth]{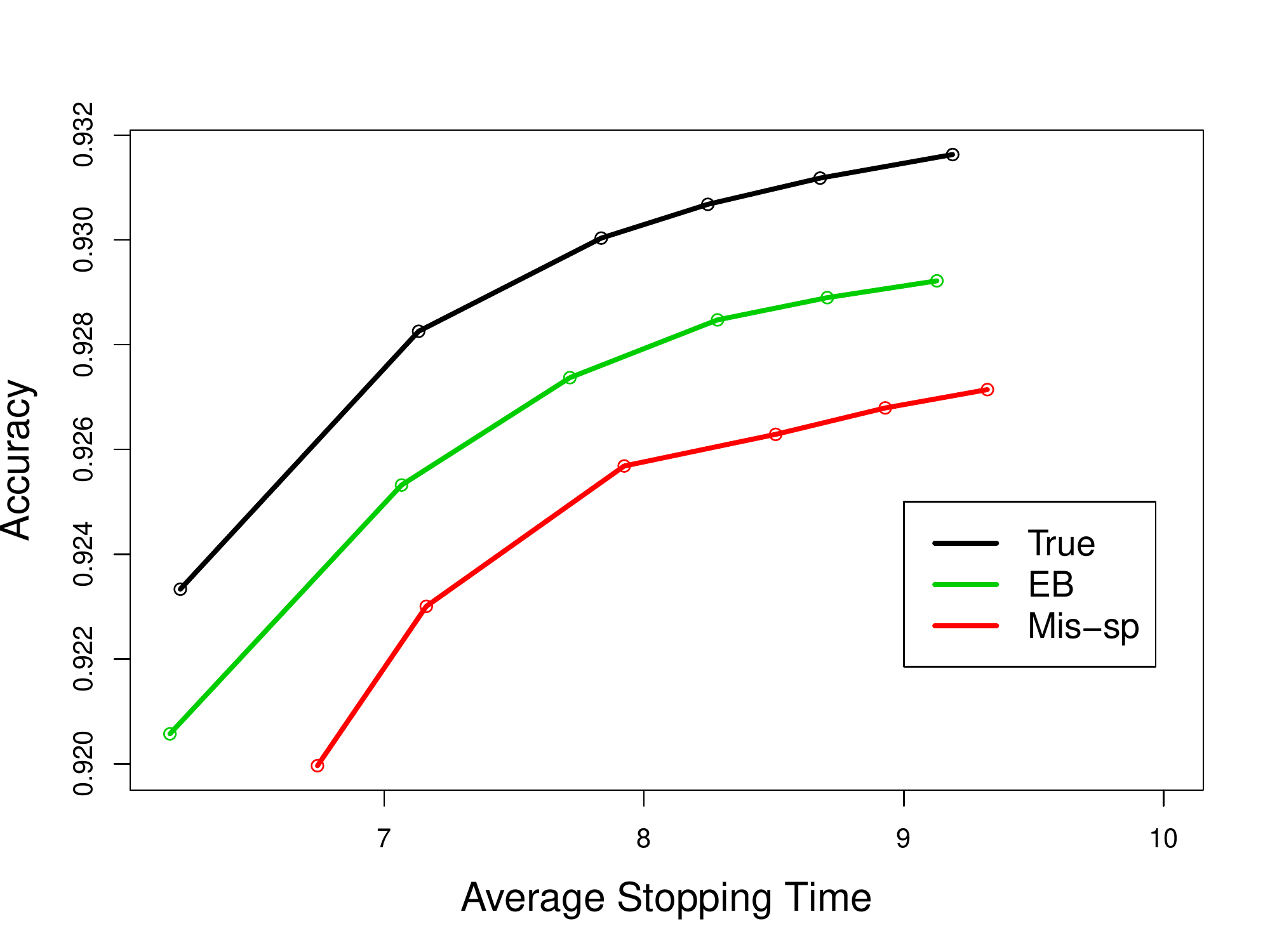}
} \vspace{0cm}
\subfigure[$\pi_1=0.5$]{
  \includegraphics[width=0.3\textwidth]{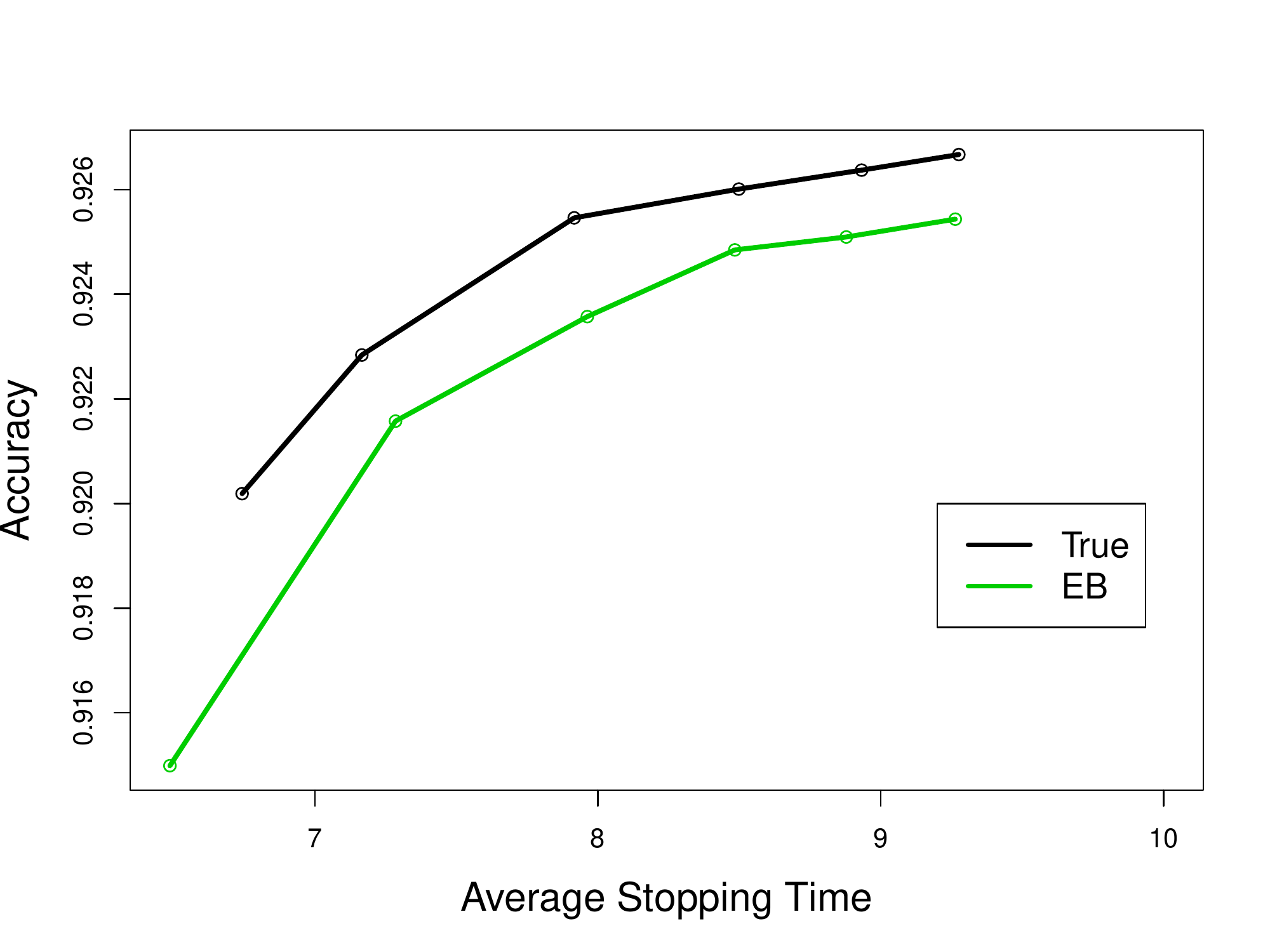}
}\vspace{0cm}
\caption{Performance of empirical Bayes estimation for different class priors.}
\label{fig:acc}
\end{figure}

In this simulated experiment, we consider the multiple hypotheses testing problem in Section \ref{sec:EB}, i.e., labeling multiple objects. In particular, we generate $K=100$ objects with true label $\theta_k$ from the Bernoulli distributions with true class prior $\pi_1$. We consider three possible values of $\pi_1$: (1) $\pi_1=0.8$ (highly unbalanced class) (2) $\pi_1=0.65$ (moderately unbalanced class) (3)  $\pi_1=0.5$ (balanced class).  For each $\pi_1$, we compare three following procedures:
\begin{enumerate}
\item  Ada-SPRT with true class prior $\pi_1$;
\item  Ada-SPRT with empirical Bayes estimation of the class prior $\pi_1$ in Algorithm \ref{alg:EMB};
\item  Ada-SPRT with the mis-specified class prior $0.5$. Note that in the third case when $\pi_1=0.5$, it is the same as the Ada-SPRT with the true class prior.
\end{enumerate}
We vary the cost parameter $c = 2^{-\rho}$ with $\rho = 7, 8, ..., 12$, which leads to different stopping times.  For each choice of $\pi_1$, we report in Figure \ref{fig:acc} the average accuracy as a function of average stopping time (i.e., $\frac{1}{K} \sum_{k=1}^K N_k$ where $N_k$ is the stopping time for the $k$-th object) for
truncated test with $T=10$ (right panels) over 5,000 independent replications. As can be seen from Figure \ref{fig:acc}, the performance of Ada-SPRT with empirical Bayes estimation is close to Ada-SPRT with true prior especially when the stopping time goes large. In addition, the performance of Ada-SPRT with empirical Bayes estimation achieves much better performance than Ada-SPRT with a mis-specified class prior, which demonstrates the effectiveness of using empirical Bayes estimation.

%
%

\subsection{Real Experiments}
\label{sec:real}
In our real experiments, we compare Ada-SPRT with several state-of-the-art worker selection approaches in crowdsourcing literature. In particular, we use two benchmark binary labeling crowdsourcing datasets:
\begin{enumerate}
  \item Recognizing textual entailment (RTE dataset \citep{Snow:08}): there are $K=800$ objects and each object is a sentence pair. Each sentence pair is presented to 10 different workers to acquire binary choices of whether the second hypothesis sentence can be inferred from the first one.   There are in total $M=164$ different workers and total number of available labels is $8,000$.  Since each object receives 10 labels, we use the truncated Ada-SPRT with the truncation length $T=10$. 
  \item Labeling bird species (Bird dataset \citep{QiangLiu:12, Peter:10}): there are $K=108$ objects and each object is an image of a bird. Each image receive 39 binary labels (either indigo bunting or blue grosbeak) from all $M=39$ workers and the total number of available labels is $4,212$. We use the truncated Ada-SPRT with the truncation length $T=39$. 
\end{enumerate}
We note that the true labels are available for both datasets from domain experts so that we could evaluate the labeling accuracy of the decision $D_k$ for each object $k \in \{1, \ldots, K\}$.
\def\arraystretch{1.2}
\begin{table}[!t]
\centering
\caption{Performance comparison on real datasets in terms of mean and standard deviation of accuracy for different approaches. \textsf{KG} and \textsf{Opt-KG} correspond to the knowledge gradient or optimistic knowledge gradient worker selection policy with the same stopping time as that of Ada-SPRT. \textsf{KG Avg} and \textsf{Opt-KG Avg} correspond to the knowledge gradient or optimistic knowledge gradient worker selection policy with the average stopping time for all objects. The accuracies in bold are the best accuracies for each choice of $c$. }
\vspace{0.2cm}
\begin{tabular}{lllll}
\hline
\textbf{RTE (Accuracy)} & $c = 2^{-6}$ &$c = 2^{-8}$ &$c = 2^{-10}$ &$c = 2^{-12}$\\  \hline
Total queried labels &3438(60) & 3949 (46)& 4365 (28)& 4660(28) \\
Ada-SPRT & \textbf{92.1\%} (0.4\%) & \textbf{92.6\%} (0.3\%) & \textbf{92.5\%} (0.3\%) & \textbf{92.6\%} (0.2\%)\\
KG       & 86.9\% (1.3\%)          & 87.4\% (0.9\%)          &  88.0\%  (1.3\%)        &  88.9\%  (1.3\%)               \\
Opt-KG   & 82.5\% (2.2\%)          & 84.3\% (2.7\%)          &  85.2\%  (1.5\%)        &  88.5\%  (1.7\%)             \\
KG Avg   & 86.1\% (2.9\%)          & 86.0\% (2.4\%)          &  87.7\%  (1.1\%)         &  87.9\%  (1.3\%)             \\
Opt-KG Avg & 82.2\% (4.3\%)        & 83.2\%  (3.2\%)                &  86.7\%  (2.0\%)               &  88.0\%  (2.2\%)              \\
\hline \hline
\textbf{Bird (Accuracy)} & $c = 2^{-6}$ &$c = 2^{-8}$ &$c = 2^{-10}$ &$c = 2^{-12}$\\
 \hline
Total queried labels & 1253 (37)& 1392 (40)& 1523 (47)& 1672 (57)\\
Ada-SPRT & \textbf{85.7}\% (4\%) & \textbf{87.5}\% (2\%) &  \textbf{87.4}\% (2\%) &  \textbf{87.1}\% (1\%)\\
KG       &  74.6\%         (5.8\%)      &  75.9\%   (3.6\%)      &  77.6\%   (4.4\%)      &   77.4\%  (3.2\%) \\
Opt-KG   &  71.3\%   (5.5\%)           &  74.4\%    (5.1\%)       & 77.1\%  (4.5\%)    &    78.1\%   (3.9\%) \\
KG Avg   &  80.4\%  (3.5\%)             &  78.8\%  (4.6\%)      &  80.0\% (2.9\%)              &    80.8\% (2.4\%)   \\
Opt-KG Avg  &  83.9\% (2.5\%)           &  84.7\%   (2.8\%)             &   85.9\% (1.8\%)             &    85.0\% (2.4\%)  \\ \hline
\end{tabular}
\label{tab:RTE}
\end{table}

For both datasets, we use truncated Ada-SPRT algorithm with EM algorithm to estimate class prior and workers' quality parameters as described in Section \ref{sec:EB}. We set $\alpha=4$ and $\beta=2$ in the regularized likelihood function in \eqref{eq:reg_likelihood}. Since $\alpha$ and $\beta$ reflect the prior belief of workers' accuracy, $\alpha=4$ and $\beta=2$ correspond to a prior accuracy of $\frac{\alpha}{\alpha+\beta}=\frac{4}{4+2}=66.7\%$. Other settings of $\alpha$ and $\beta$ lead to similar performance as long as $\alpha>\beta$ (i.e., a worker is believed to perform better than random guess).

Since different ordering of objects in Algorithm \ref{alg:EMB} leads to slightly different results, we report the average over 20 random orderings. In addition, the first quarter of the objects (i.e., the first 200 objects for RTE and the first 27 objects for Bird) will be used as a ``calibration" set. In particular, for those objects, we use all the $T$ responses (i.e., setting $N_k=T$) without selecting workers so that good initial estimates of the class prior and workers' quality parameters can be obtained based on the ``calibration'' set. \label{real:expected_stopping}  For the objects not in the ``calibration'' set, the averaged stopping times as $c$ ranging from $2^{-6}$ to $2^{-12}$ are  2.4, 3.2, 3.9, and 4.4, respectively, for the RTE dataset. For the bird dataset, the averaged stopping times as $c$ ranging from $2^{-6}$ to $2^{-12}$ are 2.5, 4.2, 5.8, and 7.6, respectively.

We compare Ada-SPRT with two state-of-the-art worker selection policies in \cite{Chen15Crowd}: (1) Knowledge gradient (KG) policy and (2) Optimistic knowledge gradient (Opt-KG) policy. We note that  both KG and Opt-KG are myopic index policy only for worker selection but not for optimal stopping. To make a fair comparison, we consider different ways of adding stopping times for KG and Opt-KG: (1) using the same stopping time $N_k$ from Ada-SPRT for each object $k$ (2) using the average stopping time $\lceil \frac{1}{K} \sum_{i=1}^K N_k \rceil$ for all objects. Recall that $N_k$ is the stopping time obtained by Ada-SPRT in Algorithm \ref{alg:EMB} for the $k$-th object. We vary the cost parameter $c$ and report mean and standard deviation of total queried labels (i.e., $\sum_{i=1}^K N_k$) and labeling accuracies  for different approaches.

The comparison results are provided in Table~\ref{tab:RTE} for RTE and Bird datasets respectively. As can be seen from Table~\ref{tab:RTE}, Ada-SPRT greatly outperforms other approaches on both datasets.  We also note that under the two-coin model, when using all the available labels, the labeling accuracy  is 92.88\% (with $8,000$ labels) for RTE dataset and 89.1\% (for  $4,212$ labels) for Bird dataset. Therefore, from Table~\ref{tab:RTE}, Ada-SPRT achieves on average $\frac{92.1}{92.88}=99\%$ of the best possible labeling accuracy using only $\frac{3438}{8000}=43\%$ of the total labels for RTE, and $\frac{85.7}{89.1}=96\%$ of  the best  possible labeling accuracy using only $\frac{1253}{4,212}=30\%$ of the total labels for Bird.

\section{Discussions}
\label{sec:conclusion}

In this paper, we propose an adaptive sequential probability ratio test (Ada-SPRT) which finds the optimal experimental selection rule,  stopping time, and  decision rule for a single hypothesis testing problem.  For multiple testing problems, we further propose an empirical Bayes approach to estimate the class prior. We demonstrate the effectiveness of our methods on real crowdsourcing applications.

There are several directions to which this work may be extended. { 
First, we only consider simple against simple hypothesis for binary labeling tasks. It is of great interest to extend the current framework to composite hypothesis. 
Second, although we mainly consider crowdsourcing applications
with a brief mention of computerized mastery testing,
 our Ada-SPRT is a general framework for adaptive sequential test, for which we would like to explore more applications.
\bibliographystyle{Chicago}
\bibliography{ref}
\clearpage
\appendix
 \begin{center}
    {\LARGE\bf Supplement to Optimal Stopping and Worker Selection in Crowdsourcing: an Adaptive Sequential Probability Ratio Test Framework}
\end{center}
  \medskip
In the supplement material, we present the proof of Proposition~\ref{thm:nontruncate}, and Theorem~\ref{thm:truncate},~\ref{thm:truncate-approximate-nontruncate} and \ref{thm:EMB}. The proof for supporting lemmas are presented in Section~\ref{sec:proof-lemmas}.

\section{Proofs of Technical Results}
\subsection{Proof of Proposition~\ref{thm:nontruncate}}\label{sec:proof_thm_non_truncate}

We consider the more general problem of finding the optimal future sequential adaptive design after collecting $n$ samples. Suppose that the first $n$ responses are $x_1,...,x_n$ and the first $n$ experiment selection functions are $j_1,...,j_n$. We need to decide the experiment selection function for the $(n+1)$'s sample, that is, $j_{n+1}(x_1,...,x_n)$. We also need to decide whether to stop the test or not and if the test is stopped, which hypothesis should be chosen. We first consider the stopping rule. To describe the stopping rule, we define the loss function
\begin{equation}\label{eq:D-loss}
L\{ (N,D),\theta \}= \ind{D\neq\theta} +c N,
\end{equation}
and the conditional risk for a test procedure $(J,N,D)$ of stopping the test with $n$ samples,
\begin{equation}\label{eq:cond-risk-n-d}
\E\Big[ L\{(N,D),\theta\}\Big|X_{1:n}=x_{1:n},N=n\Big],
\end{equation}
where we write $x_{1:n}$ as the abbreviation for the sequence $(x_1,...,x_n)$.
Because $\E\ind{D\neq\theta}= \P(\theta=0|X_{1:n}=x_{1:n})\ind{D=1}+\P(\theta=1|X_{1:n}=x_{1:n})\ind{D=0}$, it is straightforward that given $N=n$ and $X_{1:n}=x_{1:n}$, the optimal decision $D$ is
\begin{equation}\label{eq:optimal-D}
D=1 \mbox{ if } \P(\theta=1|X_{1:n}=x_{1:n})\geq \P(\theta=0|X_{1:n}=x_{1:n}) \mbox{ and } D=0  \mbox{ otherwise}.
\end{equation}
We insert this to \eqref{eq:cond-risk-n-d} and obtain the minimal conditional risk for stopping the test with $n$ samples,
\begin{align}\label{eq:conditional-risk-stopping}
r_{s}(x_{1:n},j_{1:n})
= &  \inf_{D} \E\Big[ L\{(N,D),\theta\}\Big|X_{1:n}=x_{1:n},N=n\Big] \\
= & \min\{\P(\theta=0|X_{1:n}=x_{1:n}),\P(\theta=1|X_{1:n}=x_{1:n})\}+nc. \nonumber
\end{align}
We proceed to the minimal conditional risk for continuing the test with at least $n+1$ samples,
\begin{equation}\label{eq:minimal-cond-risk}
r_{c}(x_{1:n},j_{1:n})=\inf_{(J,N,D)\in \MA_{x_{1:n},j_{1:n}}}\E\Big[ L\{(N,D),\theta\}\Big|X_{1:n}=x_{1:n}\Big],
\end{equation}
where the set $\MA_{x_{1:n},j_{1:n}}$ consists of all the sequential adaptive designs that have $j_{1:n}$ as the first $n$ experiment selection function and do not stop with $x_{1:n}$ as the first $n$ observations.


Clearly, the optimal test should continue to collect more samples if the minimal conditional risk for continuing the test is smaller than the minimal conditional risk for stopping the test. That is, the test is stopped if and only if
$$
g(x_{1:n},j_{1:n})\leq 0,
$$
where $g$ is the maximal reduced conditional risk,
\begin{equation}\label{eq:D-redueced-risk}
g(x_{1:n},j_{1:n})= r_{s}(x_{1:n},j_{1:n})-r_{c} (x_{1:n},j_{1:n}).
\end{equation}
The function $g(x_{1:n},j_{1:n})$ determines a continuing region
$
\{(X_1,...,X_n): g(X_{1:n},j_{1:n})>0\}
$
 for the sequence of samples. We further explore the shape of the continuing region. We abuse the notation a little and define the log-likelihood function
 \begin{equation}\label{eq:D-loglik-function}
 l(x_{1:n},j_{1:n})=\log\Big(\frac{\prod_{i=1}^n f_{1, \delta_i }(x_i)}{\prod_{i=1}^n f_{0, \delta_i }(x_i)}\Big),
 \end{equation}
 where $\delta_i=j_i(x_{1:i-1})$ is  the $i$-th selected experiment for $i=1,...n$.
The following lemma, whose proof is provided in Section~\ref{sec:proof-lemmas}, shows that the function $g$ depends only on the log-likelihood ratio.
\begin{lemma}\label{lemma:reduced-risk-liklihood}
There exists a function $h:\RR\to \RR$ such that for all sequence of observations $x_{1:n}$ and experiment selection functions $j_{1:n}$,
$$
g(x_{1:n},j_{1:n})=h(l(x_{1:n},j_{1:n})).
$$
\end{lemma}
According to Lemma~\ref{lemma:reduced-risk-liklihood} and the previous analysis, the optimal stopping rule is determined through the continuing region of the likelihood ratio. That is, the stopping time for the optimal design is
\[
N^*=\inf\{n: l(X_{1:n},j^*_{1:n}) \notin C \},
\]
where
\begin{equation}\label{eq:continue-region}
C= h^{-1}(0,\infty).
\end{equation}
and $j^*_{1:n}$ is the sequence of  experiment selection functions for the optimal design.

Furthermore, we describe the shape of the continuing region $C$ in the following lemma, whose proof is given in Section~\ref{sec:proof-lemmas}.
\begin{lemma}\label{lemma:shape-C}
If $a>b>\log\frac{\pi_0}{\pi_1}$ and $a\in C$, then $b\in C$. Similarly,
if $a<b<\log\frac{\pi_0}{\pi_1}$ and $a\in C$, then $b\in C$.
\end{lemma}
Lemma~\ref{lemma:shape-C} implies that the continuing region is an interval that $C=(B,A)$ for some boundary values $A$ and $B$. This completes our proof for Proposition~\ref{thm:nontruncate}(ii). In addition, we have
\begin{equation}\label{eq:theta_posterior}
\P(\theta=0|X_1,...,X_{n})=\frac{\pi_0 }{\pi_0+\pi_1 e^{l_n}} \mbox{ and }
\P(\theta=1|X_1,...,X_{n})=\frac{\pi_1 e^{l_n} }{\pi_0+\pi_1 e^{l_n}}.
\end{equation}
We insert this to \eqref{eq:optimal-D} and   Proposition~\ref{thm:nontruncate}(iii) is proved.

For the rest of the proof, we consider the optimal experiment selection.
Considering the best choice between stopping the test and continuing the test, the minimal conditional risk given the first $n$ samples $x_{1:n}$ is defined as
\begin{equation}\label{eq:minimal-condtional-risk}
U_n(x_{1:n},j_{1:n})=\min\{r_{s}(x_{1:n},j_{1:n}),r_{c}(x_{1:n},j_{1:n})\}.
\end{equation}
The optimal $(n+1)$-th experiment selection $j_{n+1}(x_{1:n})$ minimizes the future conditional risk
\begin{equation}\label{eq:D-optimal-next}
j_{n+1}(x_{1:n})=\arg\inf_{j_{n+1}(x_{1:n})}\E\Big[U_{n+1}(X_{1:n+1},j_{1:n+1})\Big|X_{1:n}=x_{1:n}\Big].
\end{equation}
Just a clarification that if the test is stopped with the first $n$ samples, then the choice of $j_{n+1}(x_{1:n})$ and does not affect the conditional risk and is thus arbitrary.
We simplify the conditional expectation in the above display
\begin{align*}
U_{n+1}(X_{1:n+1},j_{1:n+1}) = & \min\{r_{c}(X_{1:n+1},j_{1:n+1}),r_{s}(X_{1:n+1},j_{1:n+1})\}\\
                             = &  r_{s}(X_{1:n+1},j_{1:n+1})-g(X_{1:n+1},j_{1:n+1})_+,
\end{align*}
where the function $g$ is defined in \eqref{eq:D-redueced-risk} and  $x_+=\max(x,0)$.
According to Lemma~\ref{lemma:reduced-risk-liklihood} and \eqref{eq:conditional-risk-stopping}, we have
\begin{equation}
\E\Big[U_{n+1}(X_{1:n+1},j_{1:n+1})\Big|X_{1:n}=x_{1:n}\Big]=(n+1)c+ \E\Big[ u(l_{n+1})|X_{1:n}=x_{1:n}\Big],
\end{equation}
where the function $u$ is defined as
$$
u(l)=\min\{
\frac{\pi_0}{\pi_0+\pi_1 e^l},\frac{\pi_1 e^l}{\pi_0+\pi_1 e^l}
\}- h(l)_+,
$$
and $h(l)$ is defined in Lemma~\ref{lemma:reduced-risk-liklihood}.
Consequently, \eqref{eq:D-optimal-next} can be written as
\begin{eqnarray}
& j_{n+1}(x_{1:n})& \notag\\
=&\arg\inf_{j_{n+1}(x_{1:n})}  \Big\{& \P(\theta=0|X_{1:n}=x_{1:n}) \E[u(l_{n+1})|X_{1:n}=x_{1:n},\theta=0]\notag\\
&&+\P(\theta=1|X_{1:n}=x_{1:n}) \E[u(l_{n+1})|X_{1:n}=x_{1:n},\theta=1]\Big\}\label{eq:arginf-next}.
\end{eqnarray}
Notice that $l_{n+1}=l_n+\log\frac{f_{1,j_{n+1}(x_{1:n})}(X_{n+1})}{f_{0,j_{n+1}(x_{1:n})}(X_{n+1})}$ and posterior of $\theta$ is given in \eqref{eq:theta_posterior}. Therefore, \eqref{eq:arginf-next} can be written as
\begin{equation*}
j_{n+1}(x_{1:n})=\arg\inf_{j_{n+1}(x_{1:n})} v\Big(l_n, j_{n+1}(x_{1:n})\Big)
\end{equation*}
for some bivariate function $v$. Let the function
$
j^*(l)=\arg\inf_{\delta} v(l, \delta).
$
Then, we have
$
j_{n+1}(x_{1:n})=j^*(l_n),
$
and Proposition~\ref{thm:nontruncate}(i) is proved.

\subsection{Proof of Theorem~\ref{thm:truncate}}
\label{sec:proof_truncate}

Similar to the proof of Proposition~\ref{thm:nontruncate}, the stopping rule for the truncated test is determined by the maximal reduced conditional risk function
$$
g^{\dagger}(x_{1:n},j_{1:n})= r_{s}(x_{1:n},j_{1:n})-r^{\dagger}_{nc} (x_{1:n},j_{1:n}),
$$
where $r_{s}$ is defined in \eqref{eq:conditional-risk-stopping},
and $r^{\dagger}_{nc}$ is defined similarly to \eqref{eq:minimal-cond-risk},
$$
r^{\dagger}_{nc}=\inf_{(J,N,D)\in \MA^T_{x_{1:n},\delta_{1:n}}}\E\Big[ L\{(N,D),\theta\}\Big|X_{1:n}=x_{1:n}\Big]$$
and $\MA^T_{x_{1:n},j_{1:n}}$ consists of all sequential adaptive design that belongs to $\MA_{x_{1:n},j_{1:n}}$ and has a truncation length $T$. Similar to Lemma~\ref{lemma:reduced-risk-liklihood}, we establish  the following lemma, whose proof is similar to the proof of Lemma~\ref{lemma:reduced-risk-liklihood} and that of Lemma~\ref{lemma:shape-C}. 
\begin{lemma}\label{lemma:truncate-reduced-risk}
There exists a function $h^{\dagger}:\RR\times \mathbb{Z}_+\to \RR$ such that
\begin{equation}\label{eq:h-dagger}
  g^{\dagger}(x_{1:n},j_{1:n})=h^{\dagger}(l(x_{1:n},j_{1:n}),n).
\end{equation}
In addition,  for $n=1,..., T-1$, let $C_n=h(\cdot,n)^{-1}(0,+\infty)$, then we have that if $a>b>\log\frac{\pi_0}{\pi_1}$ and $a\in C_n$, then $b\in C_n$; if $a<b<\log\frac{\pi_0}{\pi_1}$ and $a\in C_n$, then $b\in C_n$. Furthermore, $ C_{n+1}\subset C_n\subset C,$ where $C$ is defined in \eqref{eq:continue-region}.
\end{lemma}
With the aid of Lemma~\ref{lemma:truncate-reduced-risk}, Theorem~\ref{thm:truncate} can be proved similarly as that of Proposition~\ref{thm:nontruncate}. We omit the details.

\subsection{Proof of Theorem~\ref{thm:truncate-approximate-nontruncate}}\label{sec:proof_approx}
For a truncated test with truncation length $T$, we consider the minimal conditional risk with $n$ samples
$$
V_n^T(x_{1:n},j_{1:n})= \inf_{(J,N,D)\in\MA^T_{x_{1:n},j_{1:n}}} E\Big[L\{(N,D),\theta\}\Big|X_{1:n}=x_{1:n}\Big].
$$
According to Lemma~\ref{lemma:truncate-reduced-risk}, $V_n^T(x_{1:n},j_{1:n})$ depends only on the log-likelihood ratio statistic $l$ that is defined in \eqref{eq:D-loglik-function}.
We abuse the notation a little and write
$$
V_n^{T}(a)= \inf_{(J,N,D)\in\MA^T_{x_{1:n},j_{1:n}}} E\Big[L\{(N,D),\theta\}\Big|l(X_{1:n},j_{1:n})=a\Big].
$$
Because $\MA^{T}_{x_{1:n},j_{1:n}}$ is increasing in $T$, so $V^{T}_n(a)$ is non-increasing in $T$ for all $n=0,1,2,...$ and $a\in\RR$.
We write
$
V^{\infty}_n(a)=\lim_{T\to\infty} V^T_n(a),
$
for each $a\in \RR$.
For each $T$, $V_n^T(a)$ follows the Bellman equation
\begin{equation}\label{eq:bellman-truncate}
V_n^T(a)=\min\Big\{ \Phi_n(a), \inf_{\delta_{n+1}} \E \left[V^T_{n+1}\left(l+\log\frac{f_{1,\delta_{n+1}}(X_{n+1})}{f_{0,\delta_{n+1}}(X_{n+1})}\right)\; \Big| \; l(X_{1:n},j_{1:n})=a \right] \Big \},
\end{equation}
where $\Phi_n(a)$ is the minimal conditional risk for stopping with $n$ samples
$$
\Phi_n(a)=\min\{\frac{\pi_0}{\pi_0+\pi_1 e^a},\frac{\pi_1 e^a}{\pi_0+\pi_1 e^a}\}+nc.
$$
Let $T\to\infty$ in \eqref{eq:bellman-truncate} and by monotone convergence theorem, we have
\begin{equation}\label{eq:inf-bellman}
V^{\infty}_n(a)=\min\Big\{\Phi_n(a), \inf_{\delta_n+1}\E\left[V^{\infty}_{n+1}\left(a+\log\frac{f_{1,\delta_{n+1}}(X_{n+1})}{f_{0,\delta_{n+1}}(X_{n+1})}\right) \vert l(X_{1:n}, j_{1:n})=a\right] \Big\}.
\end{equation}
Let $(J^*,N^*,D^*)$ be the optimal non-truncated test procedure that is defined in \eqref{eq:D-nontruncate}. According to Proposition~\ref{thm:nontruncate}, there exists experiment selection function $j^*$ such that $j^*_{n+1}(X_{1:n})=j^*(l(X_{1:n},j^*_{1:n})).$ Let $\delta^*_{n+1}=j^*(l(X_{1:n},j^*_{1:n})) $ be the stochastic process of experiment selection.
We define the following stochastic process
$$
W_{n}= V^{\infty}_n(l(X_{1:n},j^*_{1:n})).
$$
According to \eqref{eq:inf-bellman}, the process
$
\{W_n:n\geq 0\}
$
is a sub-martingale with respect to the filtration
$
\mathcal{G}_n=\sigma(l^*_m, m\leq n ),
$
where we define the stochastic process
$
l^*_m= l(X_{1:m},j^*_{1:m}).
$
To see why $\{W_n:n\geq 0\}$ is a sub-martingale,
\begin{align*}
  W_{n} = V_n^{\infty}(l^*_n) \leq & \inf_{\delta_{n+1}} \E  \left[V^{\infty}_{n+1}\left(l_n^*+\log\frac{f_{1,\delta_{n+1}}(X_{n+1})}{f_{0,\delta_{n+1}}(X_{n+1})}\right)\; \Big| \;l_n^*\right]  \\
  \leq &\E\left[ V^{\infty}_{n+1}\left(l_n^*+\log\frac{f_{1,{j}^*_{n+1}(X_{1:n})}(X_{n+1})}{f_{0,{j}^*_{n+1}(X_{1:n})}(X_{n+1})}\right)\; \Big| \;l_n^*\right] \\
  = & \E \left[ V^{\infty}_{n+1}(l_{n+1}^*)| l_n^*\right] = \E(W_{n+1} | \mathcal{G}_n).
\end{align*}
Note that $\{W_{n\wedge N^*}:n=1,2,...\}$ is uniformly integrable, where $n\wedge N^* =\min(n,N^*)$.
Using optional stopping theorem, we have
\begin{equation}\label{eq:ost}
\E[W_{N^*}]\geq W_0 =V_0^{\infty}(0).
\end{equation}
According to \eqref{eq:inf-bellman}, we have
$
W_{N^*}\leq \Phi_{N^*}(l^*_{N^*}).
$
The above display together with \eqref{eq:ost} gives
$$
\E[\Phi_{N^*}(l^*_{N^*})]\geq V_0^{\infty}(0).
$$
Note that $
\E[\Phi_{N^*}(l^*_{N^*})]= \min_{(J,N,D)\in \MA} \mathbf{R}(J,N,D)
$
and
$
V^{\infty}_0(0)=\lim_{T\to\infty}\min_{(J,N,D)\in \MA^T} \mathbf{R}(J,N,D).
$
Consequently,
\begin{equation}\label{eq:inequality1}
\lim_{T\to\infty}\min_{(J,N,D)\in \MA^T} \mathbf{R}(J,N,D)\leq \min_{(J,N,D)\in \MA} \mathbf{R}(J,N,D).
\end{equation}
The converse inequality is obvious. Since for any $T$, $\MA^T \subseteq \MA$,
\[
\min_{(J,N,D)\in \MA^T} \mathbf{R}(J,N,D) \geq \min_{(J,N,D)\in \MA} \mathbf{R}(J,N,D),
\]
which implies that,
\begin{equation}\label{eq:inequality2}
\lim_{T\to\infty}\min_{(J,N,D)\in \MA^T} \mathbf{R}(J,N,D)\geq \min_{(J,N,D)\in \MA} \mathbf{R}(J,N,D).
\end{equation}
We complete the proof by combining \eqref{eq:inequality1} and \eqref{eq:inequality2}.


\subsection{Proof of Theorem~\ref{thm:EMB}}
\label{sec:proof_EMB}
We first define the filtration $\MF_{k}$ as the $\sigma$-field generated by both the $\theta_1,...,\theta_{k}$ and the observations $X_{1,1:N_1},...,X_{k, 1:N_k}$, where $X_{k, 1:N_k}$ denotes the responses to object $k$. In addition,  let
$$
Y_k=\E\Big[L((N_k,D_k),\theta_k)|\MF_{k-1}\Big],
$$
where the loss function $L$ is defined in \eqref{eq:D-loss}.
Note that $\theta_k$ is independent with $\MF_{k-1}$. Therefore,
$$
Y_k = \tilde R(\pi_1,\hat{\pi}_1^{(k-1)}),
$$
where
\begin{eqnarray}
\tilde R(\pi_1,\hat{\pi}_1^{(k)})
&=&\pi_0 \P(D_k= 1| \hat\pi_1^{(k-1)},\theta_k=0 ) + \pi_1 \P(D_k= 0| \hat\pi_1^{(k-1)},\theta_k=1 )\notag\\
&&+c \pi_0 \E(N_k|\hat\pi_1^{(k-1)},\theta_k=0)+c\pi_1 \E(N_k|\hat\pi_1^{(k-1)},\theta_k=1).\label{eq:limit}
\end{eqnarray}
We notice the that $c\leq\hat\pi_1^{(k-1)}\leq1-c$, so the conditional expectations $\E(N_k|\hat\pi_1^{(k-1)},\theta_k=0)$ and $\E(N_k|\hat\pi_1^{(k-1)},\theta_k=1)$ are bounded.
Also notice that $\tilde R$ is a linear function in $\pi_1$ and thus Lipschitz in $\pi_1$,  so there exists a positive number $\kappa_1$ such that
\begin{equation}\label{eq:bounded}
|\tilde R(\pi_1,\hat\pi_1^{(k-1)})-\tilde R(\hat\pi_1^{(k-1)},\hat\pi_1^{(k-1)})|\leq \kappa_1 |\pi_1-\hat\pi_1^{(k-1)}|.
\end{equation}
Because $\hat{\pi}^{(k-1)}$ is consistent and \eqref{eq:bounded}, we have
$$
\tilde R(\pi_1,\hat\pi_1^{(k-1)})-\tilde R(\hat\pi_1^{(k-1)},\hat\pi_1^{(k-1)}) \xrightarrow{k \rightarrow \infty} 0 \quad\mbox{ in probability}.
$$
The next lemma shows that $\min \mathbf{R}(J,N,D)$ is also continuous in $\pi_1$. The proof for Lemma~\ref{lemma:continuous-bayes-risk} is given in Section~\ref{sec:proof-lemmas}.
\begin{lemma}\label{lemma:continuous-bayes-risk}
Let $\bar R(\pi_1)=\min \mathbf{R}(J,N,D)$ be the minimal Bayes risk corresponding to the prior probability $(1-\pi_1,\pi_1)$, then the function $\bar R(\pi_1)$ is continuous with respect to $\pi_1$. In addition, there exists a positive constant $\kappa_2$ such that for all
$c\leq\pi_1,\pi_1'\leq 1-c$
\begin{equation}\label{eq:kappa2}
|\bar R(\pi_1)- \bar R(\pi_1')|\leq \kappa_2 |\pi_1-\pi_1'|
\end{equation}
\end{lemma}
Note that $\tilde R(\hat\pi_1^{(k-1)},\hat\pi_1^{(k-1)})=\bar R(\hat\pi_1^{(k-1)})$ and $\bar R(\pi_1)=\min \mathbf{R}(J,N,D)$. By the continuity of $\bar{R}(\pi_1)$ in Lemma~\ref{lemma:continuous-bayes-risk} and the assumption $\widehat{\pi}^{(k-1)} \rightarrow \pi_1$ in probability,  we have
$$
 \tilde R(\pi_1,\hat\pi_1^{(k-1)})-\min \mathbf{R}(J,N,D) \xrightarrow{k \to \infty} 0 \quad   \mbox{ in probability}.
$$
Furthermore, according to \eqref{eq:bounded} and \eqref{eq:kappa2},
$$
|\tilde R(\pi_1,\hat\pi_1^{(k-1)})-\min \mathbf{R}(J,N,D)|\leq (\kappa_1+\kappa_2)|\hat\pi_1^{(k-1)}-\pi_1|\leq \kappa_1+\kappa_2.
$$
The above display together with the dominated convergence theorem imply that
$$
\lim_{k\to\infty}\E|\tilde R(\pi_1,\hat\pi_1^{(k-1)})-\min \mathbf{R}(J,N,D)|= 0.
$$
Consequently,
\begin{equation}\label{eq:l1bound}
\lim_{K\to\infty}\frac{1}{K}\sum_{k=1}^K \E|\tilde R(\pi_1,\hat\pi_1^{(k-1)})-\min \mathbf{R}(J,N,D)|=0.
\end{equation}
For any $\varepsilon>0$, we apply the Chebyshev's inequality and obtain
$$
\P\Big(|\frac{1}{K}\sum_{k=1}^k \tilde R(\pi_1,\hat\pi_1^{(k-1)})- \min\mathbf{R}(J,N,D)|>\varepsilon\Big)\leq \frac{1}{\varepsilon K}\sum_{k=1}^K\E | \tilde R(\pi_1,\hat\pi_1^{(k-1)})- \min\mathbf{R}(J,N,D)|.
$$
Recall $Y_k=\tilde R(\pi_1,\hat\pi_1^{(k-1)})$, then the above inequality and \eqref{eq:l1bound} give
\begin{equation}\label{eq:Y-mean-converge}
\frac{1}{K} \sum_{k=1}^{K} Y_k - \min \mathbf{R}(J,N,D) \xrightarrow{K \to \infty} 0 \quad   \mbox{ in probability}.
\end{equation}
We proceed to the limit of $L_K=\frac{1}{K}\sum_{k=1}^KL\{(N_k,D_k),\theta_k\}$. Note that
$$
\E\Big[L\{(N_k,D_k),\theta_k\}|\MF_{k-1}\Big]= Y_k.
$$
Consequently,
$
\sum_{k=1}^KL\{(N_k,D_k),\theta_k\} -Y_k
$
is a martingale with respect to the filtration $\{\MF_{K}:K=1,2,...\}$.
Standard calculation for square integrable martingale yields
$$
\E\Big[\sum_{k=1}^KL\{(N_k,D_k),\theta_k\} -Y_k\Big]^2=\sum_{k=1}^K \E[L\{(N_k,D_k),\theta_k\} -Y_k]^2\leq \kappa_3 K.
$$
for some positive constant $\kappa_3$.
We apply Chebyshev's inequality to the above display
$$
\P\Big( |L_K-\frac{1}{K}\sum_{k=1}^{K}Y_k|>\varepsilon \Big)\leq \frac{1}{K^2\varepsilon^2}\E\Big[\sum_{k=1}^KL\{(N_k,D_k),\theta_k\} -Y_k\Big]^2\leq \frac{\kappa_3}{K\varepsilon^2}
$$
for an arbitrary positive constant $\varepsilon$.
This implies  that
\begin{equation}\label{eq:LLN-martingale}
 L_K-\frac{1}{K}\sum_{k=1}^{K} Y_k \xrightarrow{K \to \infty} 0 \qquad \mbox{ in probability}.
\end{equation}
We complete the proof by combining \eqref{eq:LLN-martingale} and \eqref{eq:Y-mean-converge}.

\section{Proof of Supporting Lemmas}\label{sec:proof-lemmas}

\subsection{Proof of Lemma~\ref{lemma:reduced-risk-liklihood}}

It is sufficient to show that if
\begin{equation}\label{eq:l_equality}
l(x_{1:n},j_{1:n})=l(\bar x_{1:\bar n},\bar j_{1:\bar n}),
\end{equation}
then
$
g(x_{1:n},j_{1:n})= g(\bar x_{1:\bar n},\bar j_{1:\bar n}).
$
If in the contrary, assume without loss of generality that  $g(x_{1:n},j_{1:n})>g(\bar x_{1:\bar n}, \bar j_{1:\bar n})$, then according to the definition of $g$, there exist $(J,N,D)\in\MA_{x_{1:n},j_{1:n}}$ such that
$$
r_{s}(x_{1:n},j_{1:n})-\E^{J}\Big[L\{(N,D),\theta\}\Big|X_{1:n}=x_{1:n}\Big]>g(\bar x_{1:\bar n},\bar j_{1:\bar n}).
$$
We use the superscript $J$ in the expectation sign to indicate the expectation is computed with the experiment selection rule $J$.
We construct a sequential adaptive design $(\bar J,\bar N,\bar D)\in \MA_{\bar x_{1:\bar n}, \bar j_{1:\bar n}}$ as follows.
For any observations
$$
\bar x_{1},\bar x_2,...,\bar x_{\bar n}, y_1,y_2,....
$$
we first choose the experiment selection function $$
\bar j_{\bar n+m+1}(\bar x_{1:\bar n},y_{1:m})= j_{n+m+1}(x_{1:n},y_{1:m}).
$$
Next, for $m=1,2,...$, to decide whether the test procedure $(\bar J,\bar N,\bar D)$ stops or not with observations
$$
\bar x_1,...,\bar x_{\bar n}, y_1,...,y_m,
$$
we look at if $(J,N,D)$ stop with observations
$$
x_1,...,x_n,y_1,...,y_m
$$
or not.
If $(J,N,D)$ stops with observations $x_{1:n},y_{1:m}$ then we let $(\bar J,\bar N,\bar D)$ stop with observations $\bar x_{1:\bar n},y_{1:m}$, and otherwise we let the test $(\bar J,\bar N,\bar D)$ do not stop.
Lastly, for the decision $\bar D$ with observations $\bar x_{1:\bar n},y_{1:m}$, we also let it make the same decision as that of $D$ with observations $x_{1:n},y_{1:m}$.
In short, we let the sequential adaptive design $(\bar J,\bar N,\bar D)$ do whatever the test procedure $(J,N,D)$ do by replacing the first $\bar n$ observations with $x_{1:n}$.

We consider the reduced conditional risk for $(\bar J, \bar N,\bar D)$,
\begin{equation}\label{eq:bar-reduce}
r_{s}(x_{1:\bar n},j_{1:\bar n})-\E^{\bar J}\Big[L\{(\bar N,\bar D),\theta\}\Big|X_{1:\bar n}=\bar x_{1:\bar n}\Big].
\end{equation}
Notice that for any possible sequence of observations
$$
\bar x_{1},...,\bar x_{\bar n}, y_1,y_2,...
$$
and
$$
x_1,...,x_n, y_1,y_2,...
$$
The decision $\bar D= D$, and the stopping time
$$
\bar N-\bar n=N-n.
$$
In addition, the posterior distribution of
$X_{n+1}, X_{n+2},....$ and $ X_{\bar n+1},   X_{\bar n+2},...$ are the same with the same experiment selection rule $J$ and $\bar J$ for future experiments conditional on $X_{1:n}=x_{1:n}$ and $X_{1:\bar n}=\bar x_{1:\bar n}$ respectively.
 To see this point, notice that the conditional distribution $X_{n+1} |\theta, X_{1:\bar n}=\bar x_{1:\bar n}$ has the density function $f_{\theta, \bar{j}_{n+1}\left(\bar x_{1:\bar n}\right)}(X_{n+1})$ with the experiment selection rule $\bar J$. Since $\bar{j}_{n+1}\left(\bar x_{1:\bar n}\right)=j_{n+1}\left(x_{1:n}\right)$ by our construction, $f_{\theta, \bar{j}_{n+1}\left(\bar x_{1:\bar n}\right)}(X_{n+1}) =f_{\theta, j_{n+1}\left(x_{1:n}\right)}(X_{n+1})$, which implies that $X_{n+1} |\theta, X_{1:\bar n}=\bar x_{1:\bar n}$ has the same conditional distribution using the experiment selection rule $J$  as $X_{n+1} | \theta, X_{1:n}=x_{1:n}$. The above claim directly follows by an induction argument.
Therefore, by \eqref{eq:l_equality}, for any given $m$, we have the same conditional distribution  for the sequence $X_{n+1:n+m}|\theta,X_{1:n}=x_{1:n}$ with selection rule $\bar{J}$ and $X_{\bar n+1:\bar n+m}|\theta,X_{1:\bar n}=\bar x_{1:\bar n}$ with $J$.
Furthermore, the posterior distributions of $\theta$ are the same
given $X_{1:n}=x_{1:n}$ and $X_{1:\bar n}=\bar x_{1:\bar n}$ with selection rule $J$ and $\bar J$ respectively.
Thus, we have
\begin{equation}\label{eq:bar-risk}
\E^{\bar J}\Big[L\{(\bar N,\bar D),\theta\}\Big|X_{1:\bar n}=\bar x_{1:\bar n}\Big]-\bar n c= \E^{J}\Big[L\{( N, D),\theta\}\Big|X_{1: n}=x_{1: n}\Big]- n c.
\end{equation}
Recall that
\begin{align*}
r_{s}(x_{1:\bar n},j_{1:\bar n})=\min\{
\frac{\pi_0 }{\pi_0+\pi_1 e^{l(\bar x_{1:\bar n},j_{1:\bar n})}},\frac{\pi_1 e^{l(\bar x_{1:\bar n},j_{1:\bar n})}}{\pi_0+\pi_1 e^{l(\bar x_{1:\bar n},j_{1:\bar n})}}
\}+\bar n c,\\
r_{s}(x_{1: n},j_{1: n})=\min\{
\frac{\pi_0 }{\pi_0+\pi_1 e^{l(x_{1: n},j_{1: n})}},\frac{\pi_1 e^{l(x_{1: n},j_{1: n})}}{\pi_0+\pi_1 e^{l(x_{1: n},j_{1: n})}}
\}+ n c.
\end{align*}
Further, by \eqref{eq:l_equality},
\[
r_{s}(\bar x_{1:\bar n},j_{1:\bar n})-\bar{n} c = r_{s}(x_{1: n},j_{1: n}) - nc
\]

The above display together with \eqref{eq:bar-risk} implies
\begin{align*}
      & \; g(\bar x_{1:\bar n},\bar j_{1:\bar n}) \\
 \geq & \; r_{s}(x_{1:\bar n},j_{1:\bar n}) -  \E^{\bar J}\Big[L\{(\bar N,\bar D),\theta\}\Big|X_{1:\bar n}=\bar x_{1:\bar n}\Big] \\
 = &\; r_{s}(x_{1:n},j_{1:n})-\E^{J}\Big[L\{(N,D),\theta\}\Big|X_{1:n}=x_{1:n}\Big] \\
   >  & \; g(\bar x_{1:\bar n},\bar j_{1:\bar n})
\end{align*}
which contradicts with the assumption that $g(x_{1:n},j_{1:n})>g(\bar x_{1:\bar n}, \bar j_{1:\bar n})$.

\subsection{Proof of Lemma~\ref{lemma:shape-C}}
For $a>b>\log\frac{\pi_0}{\pi_1}$, let $(x_{1:n},j_{1:n})$ and $(\bar x_{1:\bar n},\bar j_{1:\bar n})$ be such that $l(x_{1:n},j_{1:n})= a$ and $l(\bar x_{1:\bar n},\bar j_{1:\bar n})=b$.  We assume that $g(x_{1:n},j_{1:n})>0$. For the rest of the proof, we are going to show
$$
g(\bar x_{1:\bar n},\bar j_{1:\bar n})>0.
$$
We use the similar method as in the proof of Lemma~\ref{lemma:reduced-risk-liklihood}. $g(x_{1:n},j_{1:n})>0$ implies that there exists $(J,N,D)\in \MA_{x_{1:n},j_{1:n}}$ such that
\begin{equation}\label{eq:a-reduced-risk}
r_{s}(x_{1:n},j_{1:n})-\E^{J}\Big[L\{(N,D),\theta\}\Big|X_{1:n}=x_{1:n}\Big]>0
\end{equation}
Now we construct the sequential adaptive design $(\bar J,\bar N,\bar D)\in \MA_{\bar x_{1:\bar n}, \bar j_{1:\bar n}}$ the same way as that in the proof of Lemma~\ref{lemma:reduced-risk-liklihood}. Using the same arguments as in the proof of Lemma~\ref{lemma:reduced-risk-liklihood}, we have
\begin{align}
E_0 := & \E^{J}\Big[L\{(N,D),\theta\}\Big|X_{1:n}=x_{1:n},\theta=0\Big]-nc  \nonumber \\
    =&\E^{\bar J}\Big[L\{(\bar N,\bar D),\theta\}\Big|X_{1:\bar n}=\bar x_{1:\bar n},\theta=0\Big]-\bar n c,
    \label{eq:t-0}
\end{align}
and
\begin{align}
E_1 := & \E^{J}\Big[L\{(N,D),\theta\}\Big|X_{1:n}=x_{1:n},\theta=1\Big]-nc\nonumber   \\
     = & \E^{\bar J}\Big[L\{(\bar N,\bar D),\theta\}\Big|X_{1:\bar n}=\bar x_{1:\bar n},\theta=1\Big]-\bar n c.
     \label{eq:t-1}
\end{align}
Notice that $b> \log \frac{\pi_0}{\pi_1}$ and $l(\bar x_{1:\bar n},\bar j_{1:\bar n})=b$. Consequently,
\begin{equation}\label{eq:b-risk}
r_{s}(\bar x_{1:\bar n},\bar j_{1:\bar n})= \frac{\pi_0}{\pi_0 +\pi_1 e^b}+\bar nc.
\end{equation}
We combine \eqref{eq:t-0}, \eqref{eq:t-1} and \eqref{eq:b-risk}, and arrive at
\begin{align}\label{eq:b-reduced-risk}
& r_{s}(x_{1:\bar n},j_{1:\bar n})-\E^{\bar J}\Big[L\{(\bar N,\bar D),\theta\}\Big|X_{1:\bar n}=\bar x_{1:\bar n}\Big]\\
=&\frac{\pi_0}{\pi_0 +\pi_1 e^b} - \P(\theta=0|X_{1:\bar n}=\bar x_{1:\bar n}) \times E_0 - \P(\theta=1|X_{1:\bar n}=\bar x_{1:\bar n}) \times E_1\notag\\
= & \frac{\pi_0}{\pi_0 +\pi_1 e^b}- \frac{\pi_0}{\pi_0 +\pi_1 e^b}\times E_0 - \frac{\pi_1 e^b}{\pi_0 +\pi_1 e^b}\times E_1\notag\\
=& \frac{\pi_0(1-E_0)-\pi_1 e^b E_1}{\pi_0 +\pi_1 e^b}.\notag
\end{align}
Similarly, we have
$$
r_{s}(x_{1: n},j_{1: n})-\E^{ J}\Big[L\{( N, D),\theta\}\Big|X_{1: n}= x_{1: n}\Big]=\frac{\pi_0(1-E_0)-\pi_1 e^a E_1}{\pi_0 +\pi_1 e^a}.
$$
According to \eqref{eq:a-reduced-risk} and the above display, we have
\begin{equation}\label{eq:eq1}
\frac{\pi_0(1-E_0)-\pi_1 e^a E_1}{\pi_0 +\pi_1 e^a}>0,
\end{equation}
which implies that
$$
\pi_0(1-E_0)-\pi_1 e^a E_1>0.
$$
Because $ \pi_0(1-E_0)-\pi_1 e^b E_1> \pi_0(1-E_0)-\pi_1 e^a E_1$ and \eqref{eq:eq1}, we have
$$
\frac{\pi_0(1-E_0)-\pi_1 e^b E_1}{\pi_0 +\pi_1 e^b}>0.
$$
According to the above display, the definition of $g$ and \eqref{eq:b-reduced-risk}, we have
\begin{align*}
g(\bar x_{1:\bar n},\bar j_{1:\bar n})\geq & r_{s}(x_{1:\bar n},j_{1:\bar n})-\E^{\bar J}\Big[L\{(\bar N,\bar D),\theta\}\Big|X_{1:\bar n}=\bar x_{1:\bar n}\Big] \\
\geq & \frac{\pi_0(1-E_0)-\pi_1 e^b E_1}{\pi_0 +\pi_1 e^b}>0.
\end{align*}
With similar arguments, if $a<b<\log \frac{\pi_0}{\pi_1}$ and $h(a)>0$, then we have $h(b)>0$. We omit the details.

\subsection{Proof of Lemma~\ref{lemma:truncate-reduced-risk}}
The proof of the first half of the Lemma is similar to that of Lemma~\ref{lemma:shape-C}, and is thus omitted. That is,  there exists $h^{\dagger}$ satisfying \eqref{eq:h-dagger}, and for each $C_n$, if $a>b>\log \frac{\pi_0}{\pi_1}$ and $a\in C_n$ then, $b\in C_n$. We proceed to prove that
$$
C_{n}\subset C_{n-1}.
$$
It is sufficient to show that for each $a\in C_{n+1}$, we also have $a\in C_n$. Due to the symmetry of the problem, we focus on the case where $a>\log\frac{\pi_0}{\pi_1}$.
Let $\bar n=n-1$ and let $(x_{1:n},j_{1:n})$ and $(\bar x_{1:\bar n},\bar j_{1:\bar n})$ be such that $l(x_{1:n},j_{1:n})= a$ and $l(\bar x_{1:\bar n},\bar j_{1:\bar n})=a$. We assume that $g^{\dagger}(x_{1:n},j_{1:n})>0$. For the rest of the proof, we are going to show
$$
g^{\dagger}(\bar x_{1: \bar n},\bar j_{1: \bar n})>0.
$$
We use the similar method as in the proof of Lemma~\ref{lemma:reduced-risk-liklihood}. Note that $g^{\dagger}(x_{1:n},j_{1:n})>0$ implies that there exists $(J,N,D)\in \MA^{T}_{x_{1:n},j_{1:n}}$  such that
\begin{equation}\label{eq:1}
r_{s}(x_{1:n},j_{1:n})-\E^{J}\Big[L\{(N,D),\theta\}\Big|X_{1:n}=x_{1:n}\Big]>0,
\end{equation}
where $\MA^{T}_{x_{1:n},j_{1:n}}$ is defined similar to $\MA_{x_{1:n},j_{1:n}}$ but requires that $N\leq T$.
Now we construct the sequential adaptive design $(\bar J,\bar N,\bar D)$ the same way as that in the proof of Lemma~\ref{lemma:reduced-risk-liklihood}.

Because $\bar{n}=n+1>n$, from the construction, we have
$\bar{N}=\bar{N}-\bar{n} +\bar{n}= N-n +\bar{n}=N-n+n-1=N-1\leq T$. Thus, $(\bar J,\bar N,\bar D)\in  \MA_{\bar x_{1:\bar n}^{T}, \bar j_{1:\bar n}}$.
Using the same arguments as in the proof of Lemma~\ref{lemma:shape-C},
we can see that
\begin{equation}\label{eq:2}
r_{s}(\bar x_{1:\bar n},\bar j_{1:\bar n})= \frac{\pi_0}{\pi_0 +\pi_1 e^a}+\bar nc.
\end{equation}
On the other hand, from the construction,
\begin{equation}\label{eq:3}
  \E^{J}\Big[L\{(N,D),\theta\}\Big|X_{1:n}=x_{1:n}\Big]-nc = \E^{\bar J}\Big[L\{(\bar N,\bar D),\theta\}\Big|X_{1:\bar n}=\bar x_{1:\bar n}\Big] - \bar n c.
\end{equation}
Combining \eqref{eq:1}, \eqref{eq:2} and \eqref{eq:3}, we can see that $g^{\dagger}(\bar x_{1: \bar n},\bar j_{1: \bar n})>0$. Therefore, $a\in C_{\bar n}=C_{n-1}$. This completes our proof.

\subsection{Proof of Lemma~\ref{lemma:continuous-bayes-risk}}
We consider the Bayes risk when the prior probability is $(1-\pi_1,\pi_1)$,
\begin{align*}
\mathbf{R}^{\pi_1}(J,N,D)= &
(1-\pi_1) \P(D=1|\theta=0)+\pi_1\P(D=0|\theta=1)  \\
                            &+c\{\pi_0 \E(N|\theta=0)+\pi_1\E(N|\theta=1)\}.
\end{align*}
Here we use the superscript $\pi_1$ to indicate the prior.
For fixed $(J,N,D)$ the function
$\mathbf{R}^{\pi_1}(J,N,D)$ is linear in $\pi_1$, and
is thus continuous in $\pi_1$.
Let $(J^{\pi_1},N^{\pi_1},D^{\pi_1})$ be the optimal procedure for the prior probability $\P(\theta=1)=\pi_1$. Then,
$$
\mathbf{R}^{\pi_1}(J^{\pi_1},N^{\pi_1},D^{\pi_1})=\min\mathbf{R}^{\pi_1}(J,N,D)=\bar R(\pi_1).
$$
Now we consider two prior probability $\pi_1$ and $\tilde{\pi}_1$. We have
\begin{align*}
\bar R(\pi_1)-\bar R(\tilde{\pi}_1) &  = \min\mathbf{R}^{\pi_1}(J,N,D)- \mathbf{R}^{\tilde\pi_1}(J^{\tilde\pi_1},N^{\tilde\pi_1},D^{\tilde\pi_1})
\\
& \leq \mathbf{R}^{\pi_1}(J^{\tilde\pi_1},N^{\tilde\pi_1},D^{\tilde\pi_1})
-\mathbf{R}^{\tilde\pi_1}(J^{\tilde\pi_1},N^{\tilde\pi_1},D^{\tilde\pi_1})
\end{align*}
and similarly,
$$
\bar R(\tilde{\pi}_1)-\bar R(\pi_1)\leq \mathbf{R}^{\tilde\pi_1}(J^{\pi_1},N^{\pi_1},D^{\pi_1})
-\mathbf{R}^{\pi_1}(J^{\pi_1},N^{\pi_1},D^{\pi_1}).
$$
Furthermore, for all $\pi\in[c,1-c]$ the conditional expectations $E(N^{\pi_1}|\theta=0)$ and $E(N^{\pi_1}|\theta=0)$ are bounded by some positive number $\kappa_2$.
Therefore, the continuity of $\mathbf{R}^{\pi_1}(J,N,D)$ in $\pi_1$ implies the continuity of $\bar R(\pi_1)$, and we have
$$
|\bar R(\tilde{\pi}_1)-\bar R(\pi_1)|\leq \kappa_2 |\tilde{\pi}_1-\pi_1|.
$$


\end{document}